%% file: sample-sigconf-authordraft.tex
\documentclass[acmsmall]{acmart}
\usepackage{tabularx}
\usepackage{multirow}
\usepackage{graphicx}
\usepackage{comment}
\usepackage{subcaption}
\usepackage{array}    % 支持更好的列格式控制
\usepackage[table]{xcolor}
\usepackage{booktabs} % 美观表格
\usepackage{subcaption}
\usepackage{soul}
\usepackage[utf8]{inputenc}
\usepackage{xcolor}
\usepackage[normalem]{ulem}  % 用 ulem 包处理删除线
\DeclareUnicodeCharacter{2060}{\nolinebreak}

\AtBeginDocument{%
  }

\begin{document}

\title[Moving Phones, Active Peers: Animated Phones as Facilitators in In-Person Group Discussion]{Moving Phones, Active Peers: Exploring the Effect of Animated Phones as Facilitators in In-Person Group Discussion}


\author{Ziqi Pan}
\email{zpanar@connect.ust.hk}
\affiliation{%
  \institution{The Hong Kong University of Science and Technology}
  \city{Hong Kong}
  \country{China}
}

\author{Ziqi Liu}
\email{ziqil@cs.wisc.edu}
\affiliation{%
  \institution{University of Wisconsin-Madison}
  \city{Madison}
  \state{Wisconsin}
  \country{USA}
}

\author{Jinhan Zhang}
\email{zhangjin23@mails.tsinghua.edu.cn}
\affiliation{%
  \institution{Tsinghua University}
  \city{Beijing}
  \country{China}
}

\author{Zeyu Huang}
\email{zhuangbi@connect.ust.hk}
\affiliation{%
  \institution{The Hong Kong University of Science and Technology}
  \city{Hong Kong}
  \country{China}
}

\author{Shuai Ma}
\email{mashuai@iscas.ac.cn}
\affiliation{%
  \institution{Chinese Academy of Sciences}
  \city{Beijing}
  \country{China}
}

\author{Xiaojuan Ma}
\email{mxj@cse.ust.hk}
\affiliation{%
  \institution{The Hong Kong University of Science and Technology}
  \city{Hong Kong}
  \country{China}
}

\renewcommand{\shortauthors}{Pan et al.}

\begin{abstract}
In today's in-person group discussions, smartphones are integrated as intelligent workstations; yet given their co-presence in such face-to-face interactions, whether and how they may enhance people's behavioral engagement with others remains underexplored. This work investigates how animating personal smartphones to move expressively, without compromising regular functions, can transform them into embodied facilitators for co-located group interaction. In the four-stranger small-group discussion setting, guided by Tuckman's group‑development theory, we conducted a design workshop (n=12) to identify problematic group-work circumstances and design expressive, attention-efficient animated phone facilitations. Subsequently, we developed \textit{AnimaStand}, a movement-enabled phone stand that animates phones to deliver facilitation cues according to conversation dynamics. In a between-subjects Wizard-of-Oz study (n=56) with four-stranger group discussions, where everyone's phone was on an \textit{AnimaStand}, the facilitations re-engaged inactive members, balancing interpersonal dynamics and potentially enhancing relationships, without improving task performance. We finally discuss prospects for more generalizable animated personal device facilitation.
\end{abstract}

\begin{comment}
\begin{CCSXML}
<ccs2012>
 <concept>
  <concept_id>00000000.0000000.0000000</concept_id>
  <concept_desc>Do Not Use This Code, Generate the Correct Terms for Your Paper</concept_desc>
  <concept_significance>500</concept_significance>
 </concept>
 <concept>
  <concept_id>00000000.00000000.00000000</concept_id>
  <concept_desc>Do Not Use This Code, Generate the Correct Terms for Your Paper</concept_desc>
  <concept_significance>300</concept_significance>
 </concept>
 <concept>
  <concept_id>00000000.00000000.00000000</concept_id>
  <concept_desc>Do Not Use This Code, Generate the Correct Terms for Your Paper</concept_desc>
  <concept_significance>100</concept_significance>
 </concept>
 <concept>
  <concept_id>00000000.00000000.00000000</concept_id>
  <concept_desc>Do Not Use This Code, Generate the Correct Terms for Your Paper</concept_desc>
  <concept_significance>100</concept_significance>
 </concept>
</ccs2012>
\end{CCSXML}

\ccsdesc[500]{Do Not Use This Code~Generate the Correct Terms for Your Paper}
\ccsdesc[300]{Do Not Use This Code~Generate the Correct Terms for Your Paper}
\ccsdesc{Do Not Use This Code~Generate the Correct Terms for Your Paper}
\ccsdesc[100]{Do Not Use This Code~Generate the Correct Terms for Your Paper}
\end{comment}


\begin{teaserfigure}
\centering
  \includegraphics[width=0.7\textwidth]{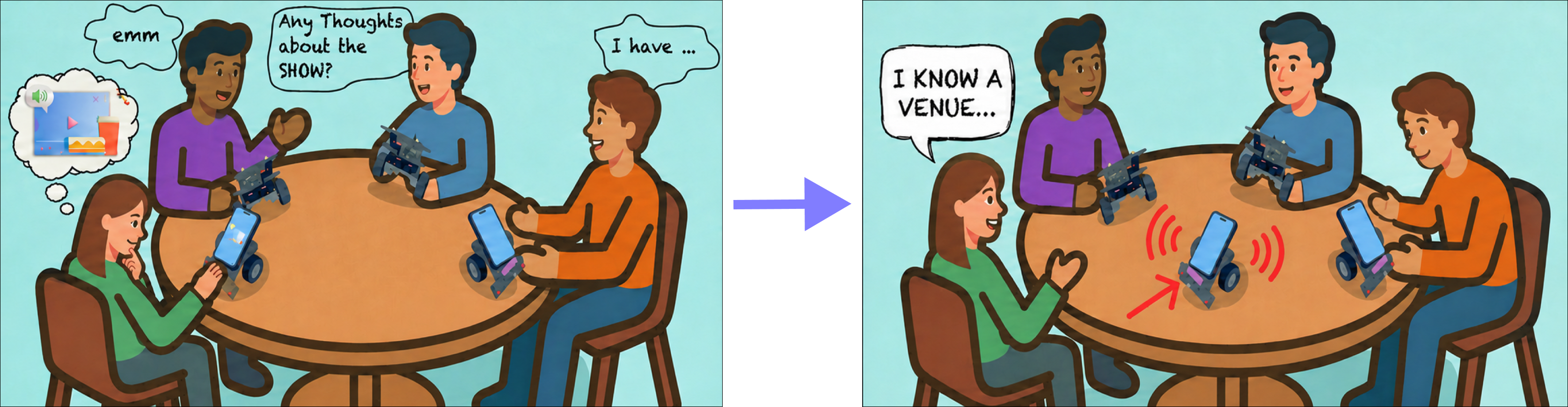}
  \caption{A typical scenario where animated phone facilitation for group discussion is helpful. When a member is zoning out, their phone steps out and dances, re-engaging them in the conversation.}
  \Description{A typical scenario where animated phone facilitation for group discussion is helpful. When a member is zoning out, their phone steps out and dances, re-engaging them in the conversation.}
  \label{fig:teaser}
\end{teaserfigure}

\maketitle

\input{1-introduction} %1163 - contribution too "exaggerated" [1k]
\input{2-related-work} %1684 - group theory too theoretical, should not emphasize [1.5k]
\input{3-design-workshop} %992 [1k]
\input{4-implementation} %902 [1k]
\input{5-method} %2470 - too much, maybe move some of the formulas to appendix? and also downgrade the role of wizard [1.5k]
\input{6-results-new} %[3k]
\input{7-discussion-new} %[1.7k]
\input{8-conclusion} %185
\bibliographystyle{ACM-Reference-Format}
\bibliography{sample-base}
%TC:ignore
\appendix
\section{Post-Experiment Questionnaire and Interview}
\begin{table}[htbp]
\centering
\caption{Post-task questionnaire and interview design}
\label{tab:questionnaire_design}
\renewcommand{\arraystretch}{0.8}
\begin{tabular}{m{3.5cm} p{7cm} p{3.5cm}}
\toprule
\textbf{Block} & \textbf{Question} & \textbf{Format} \\
\midrule
\multirow{5}{=}{Perception of \newline animated phone} 
& Q1-D1: Interpretability — ``I was able to understand what the animated phone meant.'' ~\cite{10.1145/3613905.3651027}
& \multirow{4}{*}{5-point Likert scale} \\
& Q1-D2: Influence — ``My actions were influenced by the animated phone.'' ~\cite{10.1145/3613905.3651027}
& \\
& Q1-D3: Distraction — ``The animated phone distracted our group's task completion.''~\cite{bell2023evidence}
& \\
& Q1-D4: Helpfulness — ``The animated phone helped our group's task completion.'' ~\cite{bell2023evidence}
& \\ \cmidrule{2-3}
& Q1-D5: Likeability — five dimensions \cite{bartneck2009measurement} & 5-point Likert scale \\
\midrule
\multirow{3}{*}{Perception of group work}
& Q2-D1: Confidence in the quality of the group's final plan
& \multirow{2}{*}{5-point Likert scale}\\
& Q2-D2: Satisfaction with the group's choice of plan
& \\ \cmidrule{2-3}
& Q3: Peer and self assessment — distribute 100 points among members according to contribution
& Numeric allocation (0–100) \\
\midrule
\multirow{3}{*}{Perception of relationships} 
& Q4-D1: Inclusion of the Other in the Self (IOS) scale ~\cite{gachter2025measuring}
& Visual scale (7-point) \\
& Q4-D2: We scale ~\cite{gachter2025measuring}
& 7-point scale \\
\midrule
\multirow{4}{*}{Interview}
& Q5: Reasons for Q3 score distribution  
& \multirow{4}{*}{Open-ended question} \\
& Q6: Portrait of members — brief description of each member  
& \\
& Q7: Group work experience — most unpleasant and pleasant moments, and resolutions  
& \\
& Q8: Expectations and suggestions for the animated phone facilitation  
& \\
\bottomrule
\end{tabular}
\end{table}

%TC:endignore
\end{document}

%% file: 1-introduction.tex
\section{Introduction}
% Smartphones have become a vital part of daily life. In particular, they transform productivity by offering instant and flexible access to information via searching \footnote{\url{https://www.google.com} (Google Search)}, documenting \footnote{\url{https://www.evernote.com} (Evernote)}, and sharing \footnote{\url{https://www.whatsapp.com} (WhatsApp)}. Many work-related applications and platforms on smartphones integrate these diverse functions to empower task efficiency \footnote{\url{https://slack.com} (Slack)} \footnote{\url{https://discord.com} (Discord)} \footnote{\url{https://www.microsoft.com/en-us/microsoft-teams/group-chat-software}}, expanding phones' traditional role as communication devices to mobile workstations that people constantly carry along \cite{tungare2008s}.
Smartphones have become vital companions in daily life. Serving in roles that span from communication devices to mobile productivity workstations \cite{tungare2008s}, they effectively engage users with remote collaborators and rich information in the digital space.
Yet in the physical world, given their pervasive presence in social life, whether and how smartphones may foster behavioral engagement between co-located peers remains underexplored.
Such \textbf{interpersonal behavioral engagement} is especially important in co-located group work contexts that require sustained collaborative attention and coordination \cite{10.1145/3706598.3713289}.
When this engagement drops, for example, when the group falls into pause/silence or some members free-ride with little effort, conversational flow may be disrupted~\cite{koudenburg2011disrupting} and groups' focus on collective goals may be reduced~\cite{leroy2020interruptions, hendry2016you}. These interactional issues can ultimately impair group effectiveness and rapport~\cite{walsh2019co, french1941disruption, fonti2017free}.

Previous HCI research has widely explored how co-present devices can facilitate in-person group work, specifically by fostering interpersonal behavioral engagement.
Many studies introduced external third-party artifacts, such as wearables and social robots.
Some focused on meditating socializing interactions to catalyze group relationships, utilizing physical interfaces \cite{guo2023normally, guo2025designing}, robots \cite{traeger2020vulnerable,jung2015using,10973992,zhang2023ice}
Others targeted optimizing collaborative dynamics (e.g., equity in turn-taking, participation rate) and consequently enhancing task performance through structuring collaborative behaviors in settings such as small-group conversations~\cite{zhou2026exploring}, meetings~\cite{kim2008meeting}, collaborative games~\cite{short2017robot, miura2013social}, therapeutic activities~\cite{suzuki2014social} and decision-making~\cite{10.1145/3173574.3173965}.
%Some focus on mediating relational interactions to catalyze group relationships, utilizing physical interfaces \cite{guo2023normally, guo2025designing} and robots \cite{traeger2020vulnerable,jung2015using,10973992,zhang2023ice} to encourage socializing behaviors like ice-breaking and rapport-building.
%Others targeted optimizing collaborative dynamics and consequently enhancing task performance through structuring collaborative behaviors in settings such as meetings, collaborative games, and decision-making. They employed agents or robots to enhance equity in turn-taking and participation \cite{kim2008meeting, short2017robot}, also encourage greater involvement \cite{miura2013social, suzuki2014social}. Such collaboration process facilitation improved group coordination, thereby enhancing performance \cite{10.1145/3173574.3173965}.
Moving beyond introducing new devices with independent identities, another line of work recognized the potential of leveraging intelligent devices that are already co-present with users, notably smartphones. 
These studies highlight phones' sensing, communication, and computational capabilities, equipping them with specialized apps and interfaces (e.g., AR, mobile games) \cite{10.1145/2858036.2858298, 10.1145/3706598.3713289, 10.1145/3544548.3580833} to bolster collaborative social behaviors, thus improving interactional dynamics and relationships.
However, these approaches typically organize new collaborative activities around smartphone interfaces rather than embedding facilitation into existing groupwork. Their reliance on the screen may also compete with and interrupt ongoing tasks, such as note-taking or messaging, requiring users to switch apps.
%old: However, these approaches often depend heavily on the phone's screen, which may require users to leave regular and ongoing phone usage (e.g., note-taking or messaging) to interact with a separate app.

A question thus arises: \textbf{how smartphones could become an embodied facilitator of interpersonal engagement and support co-located groups, while exercising regular productivity support without application switching}? 
To answer this question, researchers have explored leveraging smartphones' modalities beyond the primary on‑screen interface \cite{frackowiak2025co}, such as sound \cite{gencc2020designing, gencc2024designing}, to support in‑person group collaboration, typically by enabling multiple people to interact with a shared phone.
Inspired by these prior efforts, in this paper, we propose the design concept to utilize every group member's own phone, repurposing it from a passive, portable computing tool into an \textbf{actively moving device} capable of mediating behavioral engagement in co-located interactions, when necessary, through well-designed modular motions.
This design concept of making phones move actively when needed is both well-motivated and empirically supported.
First, smartphones are already perceived as functional, anthropomorphic, or even ontological extensions of the self \cite{park2019smartphone}. Thus, animating them feels less abrupt than introducing new third-party agents; rather, it is more like bringing a previously inert companion to life.
%to better support users' co-located collaboration.
Second, prior work shows that movements (e.g., proximity changes \cite{marquardt2015proxemic}, metaphorical actions \cite{urakami2023nonverbal, lakoff2008metaphors}) of smart objects, although not specifically targeting smartphones or group facilitation, can be readily perceived and interpreted as social cues and guidance \cite{10.5898/JHRI.3.1.Hoffman}.
%In this light, phones' inherent familiarity may allow users to interpret and accept their movement cues even more naturally.
%To minimize visual distraction and cognitive load \cite{simon1996designing, oulasvirta2005interaction, klein1976attention} of understanding and following animated phones' movement facilitations, we argue these movements should be attention-efficient and activated only when necessary.
% With careful design of when and how should the phones move, this approach may facilitate group work via augmenting off-screen physical affordances, while preserving the phone's primary on-screen productivity affordances.

To explore how this design concept of animated phones may facilitate groups across their lifecycle from emergence to dissolution, we conducted our study in a \textit{four‑stranger roundtable discussion} setting. 
To start with, we managed to translate our proposed design concept into a concrete design following the question: \textbf{RQ1. When and how should smartphones move to foster interpersonal engagement and facilitate collaboration in co-located groups?}
In response to \textbf{RQ1}, we first held a \textit{design workshop} (n = 12) grounded in Tuckman's group development theory \cite{tuckman1977stages}. Through iterative independent and collaborative design activities, participants identified concrete facilitation-needed circumstances, then proposed smartphone movement designs. Analyzing the resulting designs, we compiled a set of facilitation-needed circumstances with identification rules, and proposed corresponding phone movement facilitations that were expressive and attention‑efficient.
Next, to realize these movements on phones, we developed \textbf{\textit{AnimaStand}} -- a tilted phone stand with two wheels that supports moving, rotating, and signaling with lights. Rather than a standalone agent, \textit{AnimaStand} served purely as an auxiliary phone accessory to animate it.
%During group work, when certain problematic circumstances are detected automatically and confirmed manually according to the predefined rules, wizards trigger \textit{AnimaStand} to perform corresponding movements as facilitation.
During group work, \textit{AnimaStand} detected problematic circumstances from live speech diarization and recommended corresponding movement cues. For each system recommendation, a wizard checked physical feasibility, such as whether the movement path was obstructed, and then triggered the predefined facilitation movement.

Finally, we explored \textit{AnimaStand}'s effects on group collaboration in practice and how group members perceived and interpreted its facilitations. Specifically, we asked two questions:
\textbf{RQ2. How does \textit{AnimaStand} affect the processes and outcomes of small-group collaboration across interpersonal and task-activity dimensions?} and \textbf{RQ3. How do participants perceive and interpret \textit{AnimaStand} and its facilitations?}
To answer these questions, we conducted a \textit{between‑subjects experiment} (n = 56) in which 14 groups of four unacquainted participants collaborated to finish an event planning task.
Each participant placed their phone on \textit{AnimaStand}. Seven groups received animated facilitation, while seven control groups did not, with the phone stand staying still. All groups went through Tuckman's five group-development stages.
Quantitative and behavioral analyses showed that \textit{AnimaStand}'s facilitation encouraged and balanced participation and turn-taking, with participants also evaluating peers' contributions more evenly. This may be because \textit{AnimaStand}'s publicly visible and interpretable movements re-engaged inactive members back in discussion and prompted all to respond to address problematic circumstances. Facilitated groups also reported slightly higher perceived group oneness, but showed no significant task-performance improvement in objective task completion or self-assessed confidence and satisfaction (\textbf{RQ2}).
Subjective ratings and interviews further found that participants perceived \textit{AnimaStand} as generally helpful, despite occasional distraction. They made sense of it through a prediction‑and‑adaptation process with a transferring strategy, which yielded three distinct interpretive patterns without always matching the design intent (\textbf{RQ3}). 
Our work contributes to the HCI community in two ways:
\begin{itemize}
    \item We empirically explored the design of animated phone behaviors for assisting small-group, in‑person interactions, grounded in group development, proxemics, and metaphor theories.
    \item We revealed the effects of phones embodied with movements on facilitating group collaboration dynamics and outcomes regarding interpersonal and task-activity dimensions, offering implications for movement‑enabled personal devices in facilitating co‑located collaboration.
    %\item We revealed the effects of phones embodied with movements on facilitating group interactional dynamics, task operation, and interpersonal relationships. 
    %\item We investigated how group members perceive and interpret animated phone facilitations in ways that may explain their behavioral effects, thereby offering implications for movement‑enabled personal devices in co‑located collaboration.
\end{itemize}

%% file: 2-related-work.tex
\section{Related Work} % 1.5k
\subsection{Small-Group Development: Desired Goals, Challenges, and Facilitation Strategies}\label{group-dev-theory}
Small groups are fundamental collaborative units in human life, enabling members to combine diverse skills and resources toward shared objectives \cite{levine2008small}.
Prior work has extensively examined factors that influence small groups' effectiveness, including group composition \cite{belbin2022team}, cognition \cite{wegner1987transactive, cannon1993shared}, and communication \cite{gouran1996functional}, etc.
Beyond these factor-based perspectives, a family of developmental theories explains how such factors coordinate while groups develop toward effectiveness.
Tuckman's stage theory is one of the most influential frameworks from such a developmental viewpoint, initially identifying four sequential stages—\textbf{\textit{Forming}}, \textbf{\textit{Storming}}, \textbf{\textit{Norming}}, and \textbf{\textit{Performing}} \cite{tuckman1965developmental}—and later expanded to include a fifth, \textbf{\textit{Adjourning}} \cite{tuckman1977stages}. 
Following Tuckman, subsequent work has characterized stage-specific goals, challenges, and facilitation needs \cite{bonebright201040,ito2008teams,adham2023optimizing}. 
Overall, although temporary power or conversational asymmetry can also be functional (e.g., a leader or broker steering information flow)~\cite{graen1995relationship,keltner2003power}, Tuckman-inspired facilitation largely emphasizes role balance and equal voice.
In \textit{\textbf{Forming}} and \textit{\textbf{Storming}}, facilitation supports rapport and task/role clarification~\cite{ito2008teams,kim2022improving}, transparent leadership allocation, and conflict mediation ~\cite{weber1991student,sheard2004process,chou2011group,wheelan1991group,bonebright201040,ito2008teams,jones1993group}. In practice, \textit{\textbf{Norming}} and \textit{\textbf{Performing}} are often addressed jointly, with facilitation promoting cohesion, interdependence, open information flow, and balanced exchange \cite{chong2007role,jones1993group,haki2005role,sokman2023stages}. Finally, \textit{\textbf{Adjourning}} supports reflection and closure through farewells and mutual recognition of contributions \cite{kazanowski2022mixed,badyal2023small}. 

We adopt Tuckman's framework because its stage-specific goals support targeted facilitation design, while its interpersonal and task-activity dimensions guide evaluation. Applying it to short-term groups nevertheless requires adaptation. Because short-term groups may not progress through every stage, and some challenges may differ from those in established teams. For example, \textit{Adjourning} may be less consequential for one-off groups. 
To address this, taking prior work references, we propose to instantiate context-specific challenges and facilitation strategies through a design workshop, resulting in practical animated-phone facilitation designs within an adapted stage framework.

\subsection{Facilitating In-Person Group Interactions with Co-Present Devices}
Prior work in HCI has extensively examined how co‑present devices, ranging from wearables to robots to personal mobile devices, can facilitate in‑person group work.
Specifically, many works approach this facilitation by managing interpersonal behavioral engagement, as it is foundational \cite{li2013interrelations} among the three engagement dimensions (behavioral, emotional, and cognitive \cite{schaufeli2010defining, fredricks2004school}).
Earlier approaches to facilitating group interaction were rooted in group decision support systems (GDSS), in which computers moderate group discussion progress, particularly for decision making \cite{watson1988using}.
Many studies then explored introducing additional third-party mobile devices, including wearables, interfaces, and robots, into in‑person groups to enhance interpersonal relationships and collaboration dynamics. 
Focusing on cultivating the relational interaction, some leveraged physical capabilities. For example, the socio‑spatial interface \textit{SocialStool} fostered togetherness during ice-breaking through coordinated movement \cite{guo2023normally, guo2025designing, guo2025unraveling}. Others used social robots as conversation mediators to build connections, fostering trust, empathy, and positive attitudes towards group mates through approaches such as expressing vulnerability \cite{traeger2020vulnerable}, mediating conflicts \cite{jung2015using}, prompting progressive questions during icebreaking \cite{zhang2023ice}, and scaffolding emotional story exchanges \cite{10973992}.
Beyond relationship-building, other studies targeted small‑group collaboration dynamics, optimizing participation and interdependence, and finally improved task performance.
In meetings, \textit{Meeting Mediator} used real‑time speaking‑time feedback to equalize turn‑taking \cite{kim2008meeting}. In collaborative games, wearables like \textit{Enhance Reach} are illuminated to illustrate geometric distances to increase involvement \cite{miura2013social, suzuki2014social}, while moderator robots balanced participation \cite{short2017robot}. Similarly, in decision‑making tasks, embodied agents also improved both process and outcomes \cite{10.1145/3173574.3173965}.

While introducing additional devices with new identities can provide new interaction opportunities, repurposing the already co-present intelligent devices in everyday human life is also a promising and pragmatic direction.
Another line of research, in particular, explored transforming smartphones into group work facilitators. 
Although smartphones are often regarded as distractions in group settings \cite{leuppert2020commonly, barrick2022unexpected, heitmayer2021smartphones, walsh2019co}, with many studies aiming to mitigate such effects by restricting or regulating usage \cite{ko2015lock, bill2015mere, hendry2016you, ko2015nugu}, other work has emphasized their rich sensing, communication, and computational capabilities, proposing ways to harness them for group facilitation.
Research equipped smartphones with specialized agents or interfaces to facilitate various group activities. 
Examples include supporting group formation \cite{umbelino2021prototeams} and icebreaking \cite{10.1145/2858036.2858298} via mobile games, assisting co‑located team collaboration through large shared‑display interfaces \cite{10.1145/3706598.3713289}, and enabling piggybacked co‑located leisure activities with augmented reality \cite{10.1145/3544548.3580833}.
However, these solutions often rely heavily on the device's screen, which may disrupt regular functions and require interruptive app switching.
A new stream of studies thus emerged to explore leveraging smartphones' off‑screen modalities, such as sound \cite{gencc2020designing, gencc2024designing}, to support in‑person collaboration without hindering regular phone use.
Extending this direction, we investigate animating smartphones to leverage motion, operating in parallel with regular functions, to facilitate small‑group activities and enhance both collaboration dynamics and interpersonal relationships.

\subsection{Animating Devices with Proxemics and Metaphoric Movement Strategies for Collaborative Interaction Facilitation}
Animating objects has been widely studied for enhancing interaction, engagement, and collaboration \cite{ju2010animate}. This section outlines the theoretical foundations, namely proxemics and metaphoric movement theories, for designing the behavior of multiple animated personal smartphones as facilitators in small‑group collaboration, focusing on device‑scale examples rather than large‑scale installations.

\subsubsection{Proxemics}\label{proxemics}
Proxemics describes how people structure interpersonal space \cite{hall1968proxemics}. In HCI, it encompasses five dimensions---distance, identity, location, movement, and orientation---for understanding relationships among people, devices, and environments \cite{kendon1990conducting, greenberg2011proxemic, daza2021proxemic}.
Following this perspective, HCI studies mainly utilize distance and orientation as proxemic cues for social and collaborative behaviors. Closer, ``toward-and-facing'' configurations generally signal greater willingness to connect and stronger perceived relationships \cite{marquardt2015proxemic}. 
Researchers have shown that ``toward'' orientation~\cite{tatarian2021robot, connolly2021perceptions} can foster humans' inclusive behavior towards robot, while closer distance between human and devices can re-engage quieter participants and balance participation~\citet{grassi2024moderating}.
Beyond human-device proxemics, cross-device proxemics also significantly shape human-human collaboration. 
For instance, user-controlled movements of \textit{SocialStool}s helped strangers break the ice \cite{guo2023normally}, while the fixed and human-modifiable spatial arrangements of mobile devices and the environment influenced how people shared information \cite{gronbaek2020proxemics}.
Shifting the agency of changing proxemics from human control to autonomous device motions, focusing on autonomous change of the devices' orientation, \textit{MirrorBot} prompted strangers to converse when mirror-like robots were face‑to‑face \cite{guo2025mirrorbot}.

Building on this, our work explores \textbf{how autonomously moving devices drive cross-device proxemics changes to influence interpersonal behaviors}.
Particularly, we focus such exploration on personal smartphones.
Recognizing these devices as intimate extensions of the self \cite{park2019smartphone}, we investigate how their proxemic changes may serve as social cues to foster collaborative behaviors and facilitate group work.

\subsubsection{Metaphoric Movement}\label{metaphor_movement}
Metaphoric movements help express intentions of unfamiliar digital entities through familiar patterns of action. Drawing on metaphor as a structure of thought and experience \cite{lakoff2008metaphors}, prior work has embodied metaphors in digital devices' movement \cite{farnell1996metaphors, dang2025user} to support humans' interpretation and acceptance of device behaviors \cite{madill2025playing}.
At the individual level, examples include robot arms simulating nodding \cite{cao2023investigating}, pushing-like motions to attract attention \cite{kim2019swarmhaptics}, and dance-like movements inviting interaction \cite{harris2011exploring}. Such movements can encourage positive impressions and willingness to interact \cite{schulz2019animation}. At the group level, gathering, mirroring, and synchronized formations convey togetherness \cite{dang2025user,madill2025playing,kim2019swarmhaptics} and can elicit trust, positive affect, and unity \cite{kim2020conceptual,madill2025playing}.

In our study, we propose to design both \textbf{individual and group-level metaphoric movements to cope with distinct challenges throughout the collaboration}. 
Noting that movement-based facilitations may shift visual attention \cite{simon1996designing, oulasvirta2005interaction, klein1976attention}, we follow the design principle of minimizing attentional demand by using only motions that are essential to express the facilitations' intention so as to reduce unnecessary attentional load.

%% file: 3-design-workshop.tex
\section{Designing Animated Phone Behaviors to Facilitate In-Person Small-Group Collaborative Discussion}
To address RQ1, we conducted a design workshop (n=12) on animated phone facilitation in four-stranger roundtable discussions. 
We chose this configuration because newly formed groups of strangers progress through all stages of group development, and a four‑person group supports both fundamental and more complex interaction dynamics; moreover, circular seating arrangement minimizes predefined territoriality and fosters a socially inviting environment \cite{sommer1967small}.
During the workshop, participants identified circumstances requiring facilitation and designed corresponding phone movements, which were later implemented into prototype systems.

\subsection{Workshop Probes}
We designed two types of probes to prompt participants systematically identify problematic circumstances in group discussions and ideate corresponding animated phone movements as facilitations.

\subsubsection{Contextual Probe: Small-Group Development Stages, Challenges, and Goals in the Four-Person Roundtable Discussion Scenario}
To help participants systematically identify issues that may hinder collaboration across different stages of group development, we prepared a set of predefined problematic circumstances as contextual probes.
Drawing on Tuckman's theory \cite{tuckman1977stages} and related work (Sec.~\ref{group-dev-theory}), we translated stage-specific goals, challenges, and facilitation strategies into contextual probes by enumerating participant configurations in four-person roundtable discussions (Table~\ref{tab:contextual-probe}). This preliminary set was open to additions, removals, and refinements by workshop participants.

\begin{table}[!htbp]
    \centering
    \caption{Contextual probes instantiated (in italic) based on the critical analysis of group development.}
    \includegraphics[width=1\linewidth]{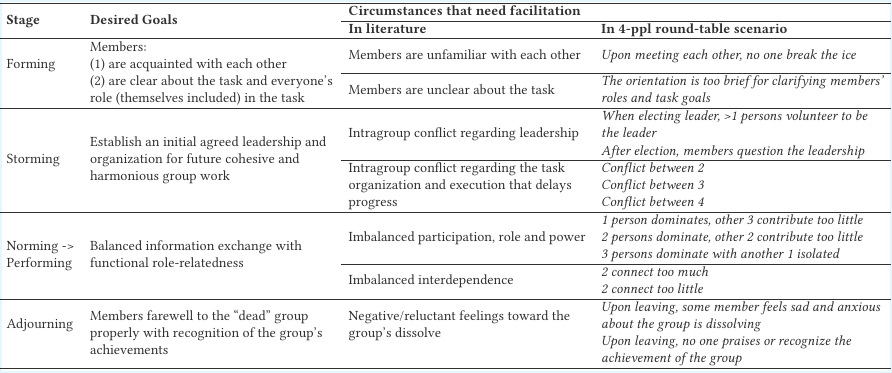}
    \label{tab:contextual-probe}
\end{table}

\subsubsection{Technical Probe: Movable Phone Stand with Controllable Lights Prototype}\label{technical-probe}
% 设计上的，大概的形态
To exemplify how a phone could be animated on a table, we built an auxiliary accessory, which is a movable phone stand (Fig.~\ref{fig:workshop}C).
This two‑wheeled phone stand with headlights holds the phone upright and slightly tilted for comfortable viewing and interaction. The stand enables forward/backward movement, rotation, and light signalling independent of motion. In the workshop, each participant received a physical prototype, manually positioning it to demonstrate plausible behaviors, and was encouraged to envision additional sensors or degrees of freedom for richer expressions.

\subsection{Workshop Procedure}
\subsubsection{Setup}
\begin{figure}[ht]
    \centering
    \includegraphics[width=0.9\linewidth]{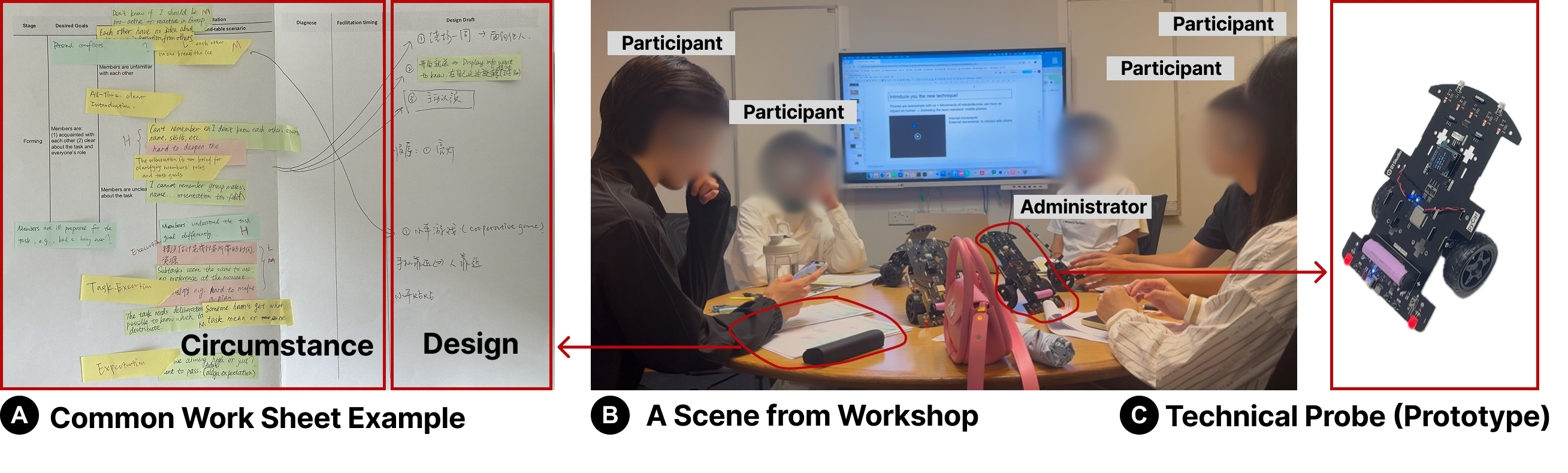}
    \caption{Demonstration of the setup of the design workshop. A. an example of the shared worksheet where participants stick their brainstorms with the administrator taking notes; B. a scene from one session when the participants were illustrating their design to the administrator; C. the technical probe, a prototype of the tilted, movement-enabled phone stand.}
    \label{fig:workshop}
\end{figure}

Each session seated four participants around a round table (Fig.~\ref{fig:workshop}B). After providing consent, participants received individual and shared worksheets, alongside phone-stand prototypes and capability cheatsheets. 
The worksheet was divided into two sections, one for documenting discussion circumstances and one for sketching corresponding design ideas, allowing participants to link situations and facilitation concepts using sticky notes and written drafts.

\subsubsection{Participants and Procedure}
With IRB approval, we recruited 12 participants (5 male, 7 female; age $23.8 \pm 4.0$) from local universities and the public, with backgrounds in Robotics, Social Science, and Design. They were mixed into three groups of four. Each session lasted approximately 120 minutes, with HKD 120 compensation and a five-minute break between activities.
After a 15-minute briefing on Tuckman's theory using video examples, participants completed two activities. In \textbf{Activity 1} (35--45 minutes), they individually revised the contextual probes, then consolidated and ranked circumstances by impact and urgency\footnote{Impact-Urgency-Matrix \url{https://miro.com/strategic-planning/impact-urgency-matrix/}} together. In \textbf{Activity 2} (35--45 minutes), they individually proposed movements for prioritized circumstances using text, sketches, or videos, then jointly refined and expanded their designs. Finally, participants demonstrated viable designs through four-person role-play with the phone-stand prototypes.

\subsection{Data Analysis}
Using the workshop recordings and worksheets, two authors conducted a theory‑driven deductive–inductive analysis of participants' ideation on problematic group discussion circumstances and animated phone behaviors. We first refined the set of problematic circumstances by discarding contextual probe entries without participant-proposed facilitation designs and merging participant‑suggested edits from \textit{Activity 1}. For the corresponding phone animation designs from \textit{Activity 2}, we deduplicated ideas and analyzed the resulting concepts through two lenses. First, following \citet{lakoff2008metaphors}, we identified each design's underlying interaction intention (e.g., expressing closeness, signaling turn‑taking). Second, using an extended proxemics framework (Sec. \ref{proxemics}), we coded designs along five socio‑spatial dimensions (distance, orientation, movement, position, identity) to surface recurring patterns. Finally, drawing on work on attention as a limited resource \cite{simon1996designing, oulasvirta2005interaction}, we synthesized designs for each refined circumstance that balance expressiveness and attention economy by assigning only the most relevant phones the minimal actions needed to achieve intended interaction goals.

\subsection{Findings}
\subsubsection{Circumstances Require Facilitation Modified by the Participants}
Analyzing participants' feedback on the circumstance probes, we identified two cross‑group modifications.
First, all groups recommended \textbf{adding ``All Members Being Silent''} to both the \textit{Storming} and \textit{Norming–Performing} stages, viewing it as a more urgent concern than leadership competition during the \textit{Storming} stage. Two groups (G1 and G2) further noted that, in the \textit{Norming–Performing} stage, members may also work solely without collaborating. This emphasis may stem from the workshop's focus on collaboration among strangers rather than more established teams in literatures.
Second, participants recommended \textbf{adding ``Conflict between Members'' also to the \textit{Norming-Performing} stage}, arguing that task‑focused disagreements are more common than leadership disagreements in a stranger group, and likely to intensify as work progresses.
Although expert-led task completion can be effective in some contexts, participants prioritized equal contribution, characterizing effective groups as combining task efficiency with a positive interpersonal climate. This preference was likely shaped by our four-stranger setup, which did not presuppose ability differences, and by the participant pool: many were university students accustomed to assessment contexts (e.g., graded group projects) that emphasize equal contribution.

\subsubsection{Identification of the Problematic Circumstances}
Participants also provided insights on how to identify the problematic circumstances.
They generally proposed an approach that combines \textbf{conversational features with context-aware interpretation}. 
Consistent with our earlier findings of desired facilitation goal, they implied participation and interdependence as key indicators of group dynamics in their suggestions of concrete detection cues.
For example, extended speech or prolonged silence could signal excessive or insufficient participation (G1, G3). While frequent two-person turn-taking might indicate interdependence, when combined with contextual cues such as volume level and discussion topic, it can be interpreted as a sign of conflict (G2). 
Participants also noted that \textbf{while some situations should be detected automatically, others may be better left for users to assess directly}. For instance, it would be better to let users decide whether they lack connection with another member (e.g., upstream/downstream members), based on the task's progress, rather than relying on automatic detection (G2, G3).
Moreover, participants noted that \textbf{some circumstances that often arise when entering a new stage are not necessarily problematic, but still worth supporting}.
For example, in the adjourning stage, when the group is about to disband, although failing to conclude properly may have little impact on group performance as the group no longer exists, ending with a friendly exchange of contact information and goodbyes can still be valuable (G1, G2).

\subsubsection{Movement Design Ideations}

\textbf{The role of the animated phone facilitation differs in different cases.}
Participants' designs assigned two types of roles to the animated phones based on members' awareness of the collaborative issue and their intent to intervene.
When members recognized an issue but felt socially constrained, phones acted as \textit{spokespeople}, for example by approaching or signaling another phone to request attention without interrupting (G1, G2, G3). When members were unaware of an issue, phones acted as \textit{role models}, demonstrating desired behavior. For example, a silent member's phone could step out and rotate to invite participation from its owner and peers.
\textbf{Design animated phone facilitations in modules.}
Across groups, participants repeatedly reused similar animations to express similar facilitation intentions, suggesting a modular design approach. 
Both the formation of a group of phones and the individual movements could be reused for different circumstances across stages. Group formations like phones gathering and moving synchronously could be leveraged to support cohesion in orientation, leader election, and adjourning. While individual movements could be mapped to human-like behaviors in the same way across circumstances, for instance, blinking lights to indicate speaking, moving to the center and rotating to attract attention, or orienting toward another phone to convey support. For example, in the cases of 1, 2, or 3 members dominating, G1 proposed one design where less active members' phones move to the center and rotate to attract attention, adjusting only the number of rotating phones to match each case. Such modularity supports coherent, interpretable facilitation with lower cognitive load across circumstances.

\begin{figure}[htbp]
    \centering
    \includegraphics[width=0.9\linewidth]{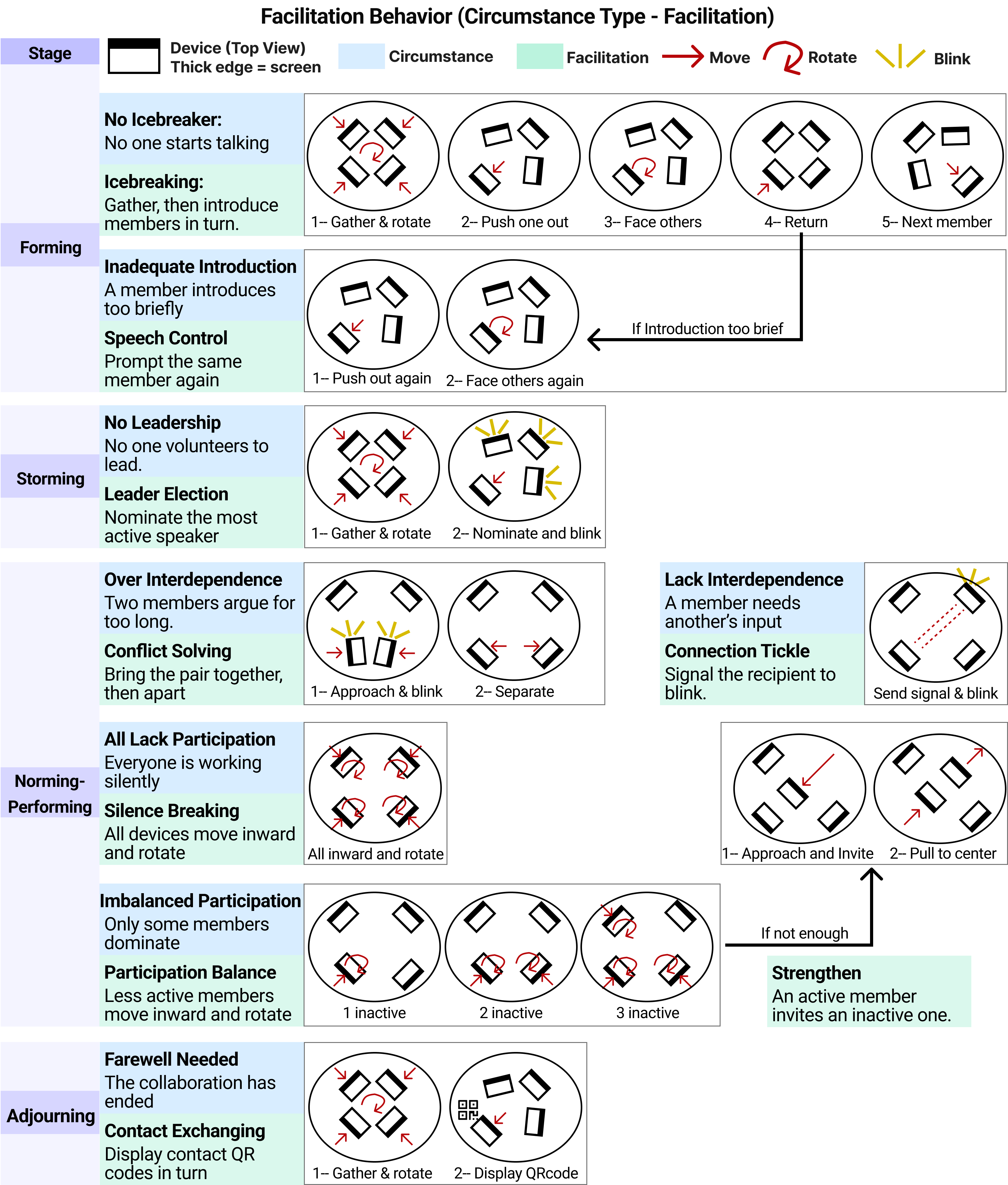}
    \caption{Final design of animated phone facilitation behaviors across small-group development stages, showing circumstance types with qualitative identification rules, and facilitation designs with brief descriptions and figure demonstrations.}
    \Description{The figure summarizes how AnimaStand's phone behaviors map onto Tuckman's stages and typical group circumstances. In the Forming stage, phones act as icebreakers: when no one starts talking, they gather and rotate in synchrony, then sequentially push one phone outward to cue self‑introductions; phones that gave too little information are pushed out again and turned toward others as a prompt to elaborate. In the Storming stage, phones help structure leadership and mitigate conflict: when there is no leader, all phones gather and rotate, then one phone is pushed forward while others blink to indicate agreement; when two members quarrel, their phones approach and blink to acknowledge the dispute, then dismiss and separate to signal resolution, and ``tickling'' motions invite specific partners when interdependence is needed. In the Norming–Performing stage, phones support participation balance and re‑engagement: when everyone has gone quiet, all phones step toward the center and rotate to suggest group discussion; if only some members dominate, inactive members' phones step forward and rotate, and—when needed—an active member's phone approaches and pulls an inactive one toward the center. In the Adjourning stage, phones facilitate closure by gathering and rotating together, then sequentially facing the group while showing their owners' contact QR codes to prompt farewell and contact exchange.}
    \label{fig:final-design}
\end{figure}

\subsection{Final Animation Design}\label{final-design}
Based on the findings, we finalized the circumstances that require facilitation, along with the qualitative identification rules and corresponding facilitation design, as presented in Fig. \ref{fig:final-design}.

%% file: 4-implementation.tex
\section{Implementation}
\begin{figure}[htbp]
    \centering
    \begin{subfigure}[b]{\textwidth}
    \includegraphics[width=0.9\linewidth]{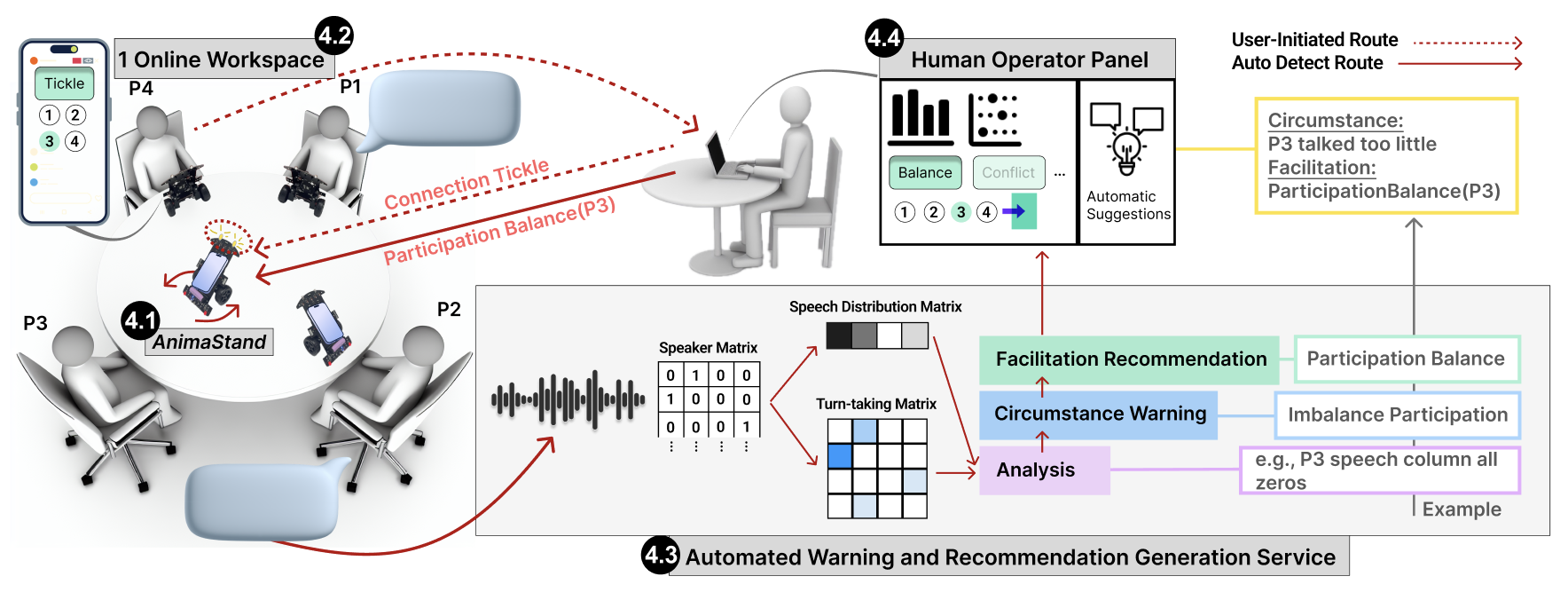}
    \caption{Overall workflow of the \textit{AnimaStand} system, illustrating data flow of two types: (1, solid line) from live speech processing to recommendation generation, human administrator mediation, and animated facilitation execution; (2, dashed line) from user-initiated command from their Slack workspace, to server, and another user's \textit{AnimaStand}}
    \Description{}
    \label{fig:overall}
    \end{subfigure}
    \begin{subfigure}[b]{0.42\textwidth}
        \centering
        \includegraphics[width=\linewidth]{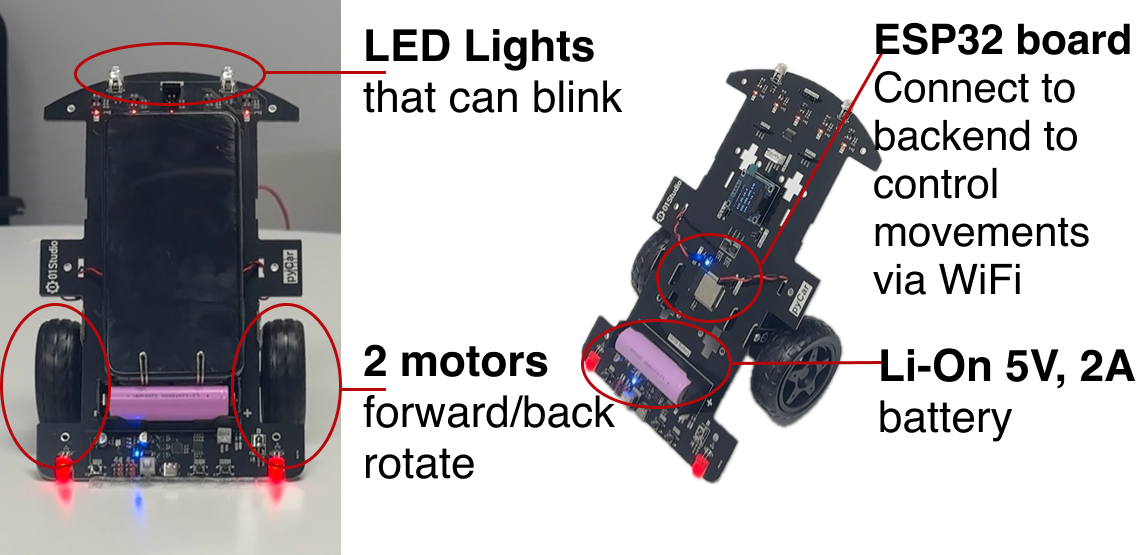}
        \caption{\textit{AnimaStand} prototype.}
        \label{fig:animastand}
    \end{subfigure}
    \begin{subfigure}[b]{0.42\textwidth}
        \includegraphics[width=\linewidth]{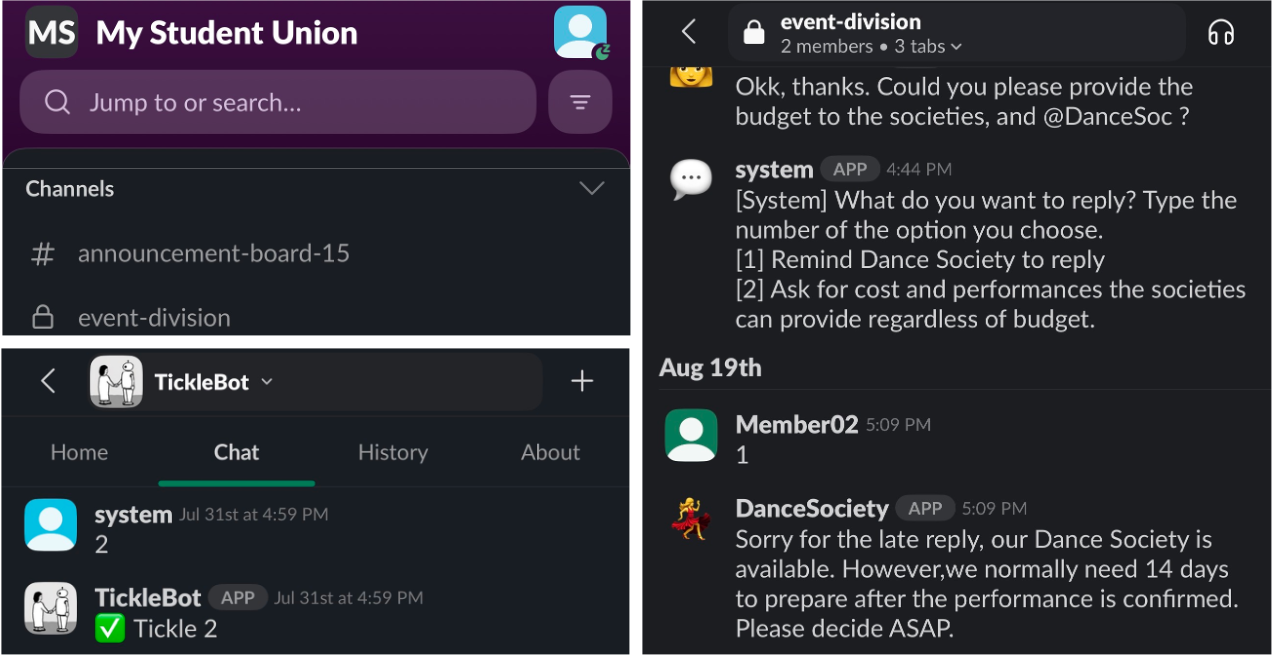}
        \caption{Online Slack workspace. Left top: the channel page; left bottom: the Tickle bot example; right: one private channel example.}
        \label{fig:slack}
    \end{subfigure}
    \caption{Overall workflow and demonstration of \textit{AnimaStand} with online Slack workspace.}
\end{figure}

Building on the design workshop outcomes, we developed a facilitation system for our study to explore the effects of animated phone facilitation on group work, while maintaining the devices' regular communication and productivity functions.
The facilitation system comprises four elements: motorized \textit{AnimaStand} phone stands, a Slack workspace\footnote{Slack Technologies LLC, https://slack.com}, a speech-analysis backend, and an administrator control panel (Fig.~\ref{fig:overall}). Slack simulates task-related productivity usage of phones and supports user-initiated \textit{Connection-Tickle} requests.
The backend recommends facilitations based on live conversational dynamics. An administrator reviews these recommendations through the control panel, checks physical feasibility (e.g., whether objects obstruct the phones' movement paths), and triggers the movements as soon as possible. The remainder of this section details the system's four primary modules and their workflow.

\subsection{Hardware implementation of \textbf{\textit{AnimaStand}}}\label{animastand-intro}
The \textit{AnimaStand} (Fig.~\ref{fig:animastand} with technical details) is a mobile phone stand with a tilted backplate and adjustable clamps for securing the phone at an optimal viewing angle.
Rather than a standalone agent, \textit{AnimaStand} served purely as an auxiliary accessory to animate the phone.
It can perform forward/backward motion, in‑place (counter)clockwise turning with two controllable LED headlights that can blink to emit signals. 
Handled by an ESP32 development board and connected to the back‑end control system via Wi‑Fi, \textit{AnimaStand} can be operated to perform the modular movements when needed as designed in Fig.~\ref{fig:final-design}.
%It features two independently driven motors, each capable of forward and reverse rotation, enabling linear motion and in-place turning, with an optical encoder for wheel‑rotation measurement, distance calibration, and precise movement control. Two controllable LED headlights are integrated into the backplate for signalling.
%Motor and light control are handled by an ESP32 development board, offering ample GPIO support, low‑latency response, and Wi‑Fi connectivity. The stand communicates with the back‑end control system via Wi‑Fi using the MQTT protocol. The \textit{AnimaStand} as a whole is powered by a rechargeable Li‑ion battery pack (5V, 2A output), providing mobile operation for up to 5 to 6 hours.

\subsection{Online Slack Workspace}\label{slack}
The online Slack workspace (Fig.~\ref{fig:slack}) comprises a general channel, private channels, and a \textit{TickleBot}. 
The general and private channels create a realistic scenario where members need to integrate their exclusive knowledge to complete a common task.
In particular, the general channel functions as the group's shared information space, where members receive collective task announcements and post self-introductions.
In contrast, the private channels are exclusively assigned to a single member and simulate the information asymmetry in real-world group work. In each private channel, members need to search pre-configured chat history and query about role-specific information (e.g., budget constraints for some roles) with an auto-reply chatbot.
Finally, the \textit{TickleBot}, a Slack bot that enables silent connection requests, allowed members to discreetly notify others without verbally interrupting ongoing discussions (user-initiated \textit{Connection-Tickle} in Fig. \ref{fig:final-design}); When a member sends a ``\textit{tickle}'' through the bot, the target member's \textit{AnimaStand} blinks its LED lights as a visual cue, signaling that someone wishes to get their attention.

\subsection{Automated Warning and Facilitation Recommendation Generation for Problematic Circumstances}

\subsubsection{Live Diarization and Speech Data Processing} \label{4.3.1}
The back‑end web application first performs live speaker diarization on the audio stream captured during discussions. 
Based on all participants' pre‑registered voices before each session, we employ the Pyannote toolkit to identify the primary speaker of each second. The diarization did not include the Auto Speech Recognition (ASR)-based content extraction, as ASR accuracy issues could introduce systematic biases.
% We employ the Pyannote toolkit for diarization \cite{Plaquet23, Bredin23}, with all participants pre‑registered before each session, enabling the system to identify the primary speaker in every one‑second interval. 
The diarization output is stored as a growing $ n \times 4$ matrix, where each column corresponds to a participant and each row indicates their speaking activity at a given second-level timestamp.
For each current timestamp $t$, we apply a sliding‑window analysis (defined as one \textit{evaluation}) over the preceding 60 seconds $[max(0, t-60s), t]$ to compute two conversation feature matrices. These matrices provide a real‑time quantitative view of participation and interdependence dynamics, which is key to subsequent detection of problematic circumstances.
\begin{itemize}
    \item \textbf{Speech distribution matrix}: a $1 \times 4$ matrix recording the cumulative speaking time of each participant within the window. An unsupervised k-means clustering algorithm ($k=2$) is then applied to automatically distinguish between dominant and non-dominant speakers based on their relative speaking times.
    \item \textbf{Directed turn‑taking matrix}: a $4 \times 4$ matrix in which each entry $(i, j)$ records the number of times speaker $i$ is followed by speaker $j$, capturing the turn transitions with directions.
\end{itemize}
We selected a 60‑second window based on pilot testing and prior work showing that the typical duration per speaker per turn is 15–20 seconds. In four‑person groups, this 60-second window accommodates roughly one turn per member. 
% Moreover, as confirmed in pilot testing, windows under 45 seconds often missed trends, and over 75 seconds delayed facilitation.
%These matrices provide a real‑time quantitative view of participation and interdependence dynamics, which serve as key inputs to the subsequent detection of problematic circumstances in group collaboration.

\subsubsection{Warning and Recommendation Generation Algorithm}\label{algorithm}
\begin{figure}[htbp]
    \centering
    \includegraphics[width=0.9\linewidth]{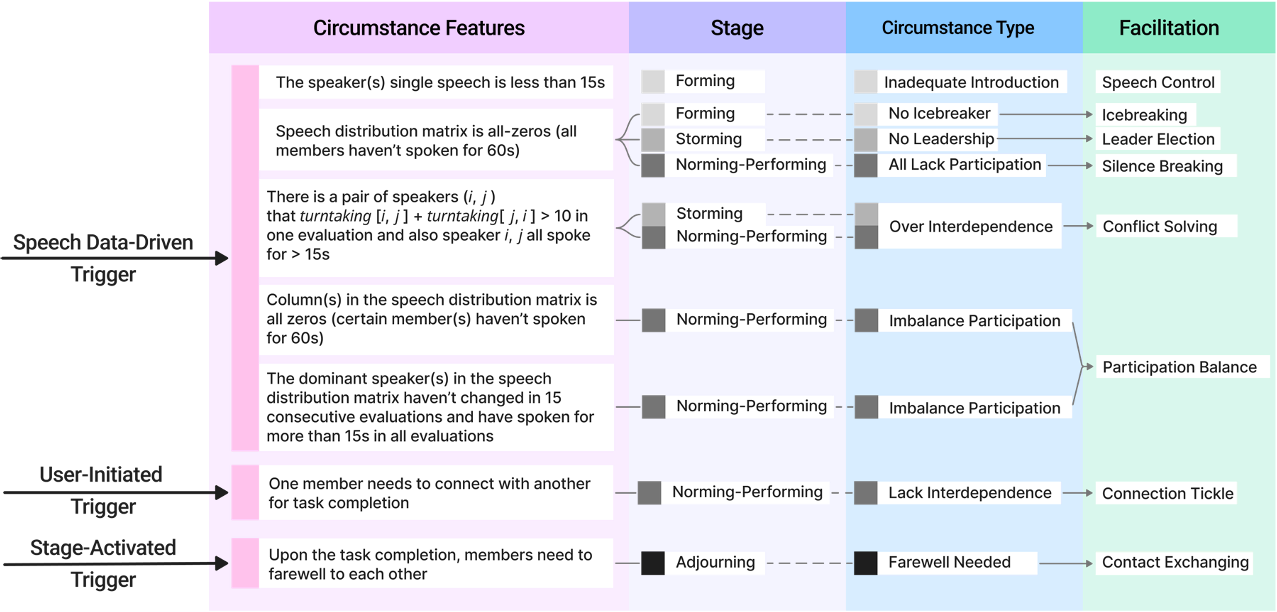}
    \caption{Implementation mapping from circumstance features to group development stages, circumstance types, and corresponding animated phone facilitation designs.}
    \Description{}
    \label{fig:manual-book}
\end{figure}

Leveraging the above metrics, we implemented rule‑based detection of problematic circumstances as shown in Fig. \ref{fig:manual-book}.
Because workshop participants described how to identify problematic circumstances qualitatively (e.g., ''a certain amount of time''), we established our final time thresholds and parameters by calibrating empirical speaking-turn baselines (15–20 seconds \cite{sacks1974simplest}) through pilot trials.
The resulting recommendations were not executed automatically because the system could not detect physical obstacles. Before execution, a human administrator only checked whether the phone's movement path was clear and triggered the recommended facilitation as soon as it was physically feasible; no conversational or semantic judgment was involved.
%In our design workshop, participants described how to identify the problematic circumstances qualitatively (e.g., “a certain amount of time”), without specifying exact thresholds. 
%Therefore, we referred to prior studies, which indicated that typical speaking turns in multi-party discussions last 15–20 seconds empirically \cite{sacks1974simplest}, and conducted pilot trials to calibrate parameters for our context, resulting in the final thresholds used in our rules.
% These rules serve only as warnings rather than automatic triggers, because our live diarization data did not capture enough contextual information, as it did not extract content to avoid ASR accuracy issues. Because our live diarization only captured speaker segmentation and voiceprint features without richer content extraction to avoid ASR accuracy issues that could introduce systematic biases.
%relies solely on speaker segmentation and voiceprint features, without Auto Speech Recognition (ASR)-based content extraction, to avoid ASR accuracy issues that could introduce systematic biases. 

\subsection{Control Panel}\label{control-panel}
\begin{figure}[h]
    \centering
    \includegraphics[width=0.85\linewidth]{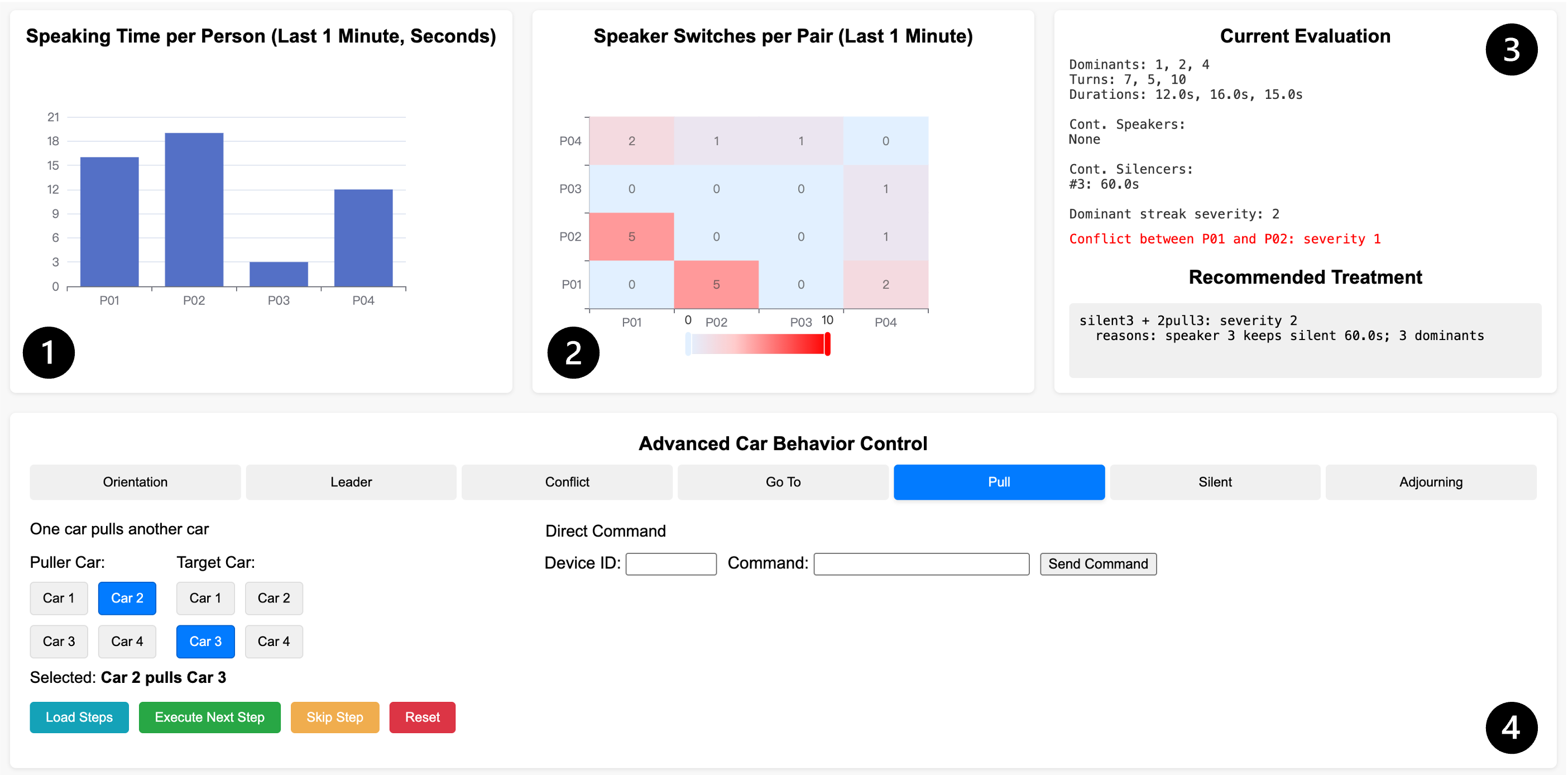}
    \caption{Control Panel. The panel consists of four main blocks: 
    (1) real-time speaking time per participant, 
    (2) speaker switches per pair in the last minute (the sum of two directions), 
    (3) current evaluation with detected issues and recommended treatments, 
    and (4) advanced control buttons for triggering facilitation behaviors. 
    Together, the administrator used the control panel to monitor system recommendations, check physical feasibility, and trigger predefined facilitation movements.}
    \label{fig:controller-panel}
\end{figure}

The control panel displays live speech and turn-taking matrices alongside warnings and facilitation recommendations (Fig.~\ref{fig:controller-panel}). Upon receiving a recommendation, the panel displayed the predefined facilitation and its targets. The administrator only checked physical feasibility and triggered the movement sequence once the path was clear. Direct controls are available for unexpected situations.

%% file: 5-method.tex
% TODO: condense and move some part to the appendix
\section{Method: Exploring Animated Phone Facilitation in In-Person Small-Group Discussions}
Using the \textit{AnimaStand} and its integrated facilitation system, we conducted a between‑subjects Wizard‑of‑Oz experiment to examine the effects of animated phone‑based facilitation on group processes (RQ2) across Tuckman's group development stages—\textit{Forming}, \textit{Storming}, \textit{Norming‑Performing} (combined as justified in Sec. ~\ref{group-dev-theory}), and \textit{Adjourning}. Post-experiment questionnaires and interviews further investigated how users' understanding of this approach accounts for the effects (RQ3).

\subsection{Task Design}\label{task-design}
\begin{figure}[ht]
    \centering
    \includegraphics[width=\linewidth]{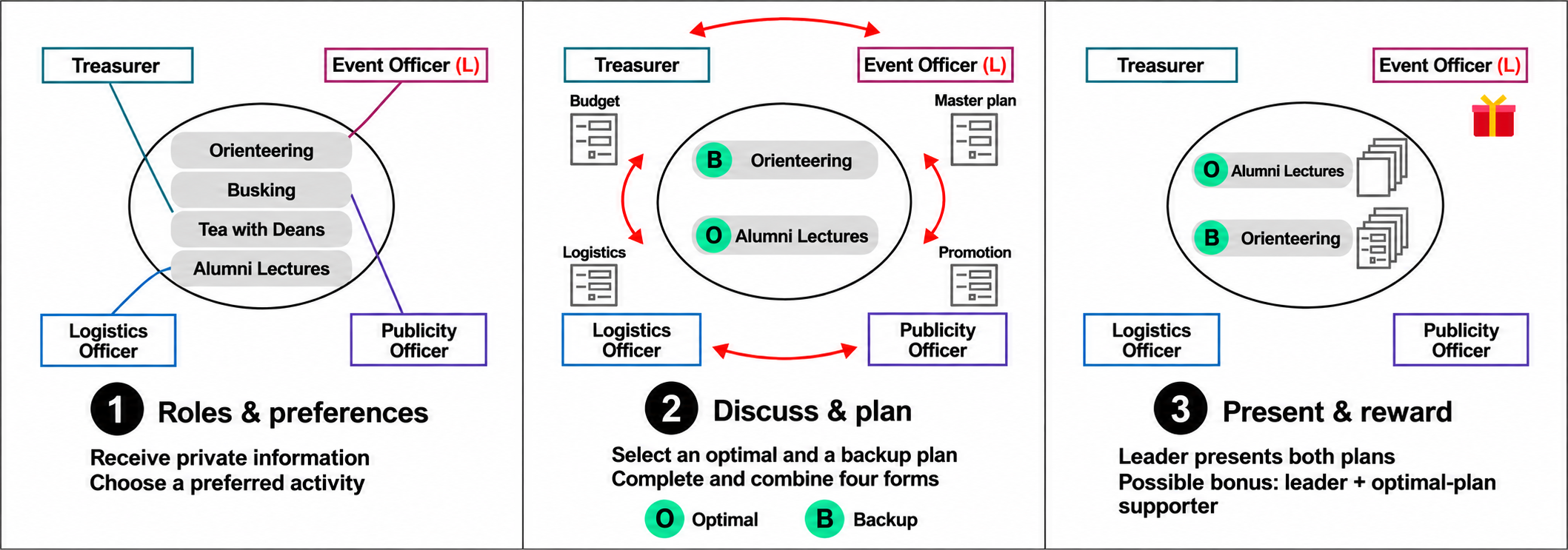}
    \caption{Illustration of the task design and procedure. 
    (1) Initial status after role election, with role-exclusive information on Slack and paper materials, and preference for a particular proposal;
    (2) Task completion procedure;
    (3) Expected outcomes and bonus scheme.}
    \Description{}
    \label{fig:task-design}
\end{figure}

We designed a four‑person role‑play task to examine the influence of animated phone facilitation during the \textit{Storming} and \textit{Norming–Performing} stages, with the \textit{Forming} and \textit{Adjourning} stages examined in the pre‑ and post‑task procedures. In the task, participants acted as officers from four different student‑union divisions collaboratively planning a Freshers' week orientation event (as shown in Fig.~\ref{fig:task-design}). 
The task comprised role and leader election (\textit{Storming}), selecting an optimal and a backup plan from four proposals, and completing role-specific preparation sheets for both plans (\textit{Norming--Performing}). Participants can choose their own task-completion workflow. 
The task should possibly evoke the problematic circumstances in group work as listed in Fig. \ref{fig:final-design}, so that facilitations could be triggered and sequentially evaluated. While challenges of \textit{Forming} and \textit{Adjourning} stage naturally occur for our four-stranger new group setting, four task features were designed to elicit leadership competition, conflict, participation, and interdependence challenges during \textit{Storming} and \textit{Norming-Performing} (Fig.~\ref{fig:final-design}):

\begin{itemize}
    \item \textbf{Leadership competition:} Leaders had greater workloads and an additional HKD 20 bonus for completing over 85\% of their departmental sheet for the optimal plan.
    \item \textbf{Conflict of interest:} Each participant had a private, role-specific plan preference and a HKD 20 bonus if that plan was selected.
    \item \textbf{Participation:} Event, Treasurer, Logistics, and Publicity Officers received exclusive information through private Slack channels and departmental materials, requiring contributions from all members.
    \item \textbf{Interdependence:} Each departmental sheet required information held by other members, necessitating dyadic exchange.
\end{itemize}

\subsection{Experiment Procedure}
\subsubsection{Participants}
Approved by IRB, we recruited 56 participants through social media and word-of-mouth (21 male, 35 female, Age = $24.2 \pm 2.9$). 
Participants were then assigned to 14 separate study sessions (seven with \textit{AnimaStand}s' facilitation, seven without), each comprising four unacquainted individuals. 
Participants reported their prior experience with teamwork, intelligent assistants, and online productivity tools during recruitment. These responses were used to allocate participants to sessions to balance comparable familiarity with the experiment task background across groups.

\subsubsection{Environment Setup}
\begin{figure}
    \centering
    \begin{subfigure}[b]{0.188\textwidth}
    \includegraphics[width=\linewidth]{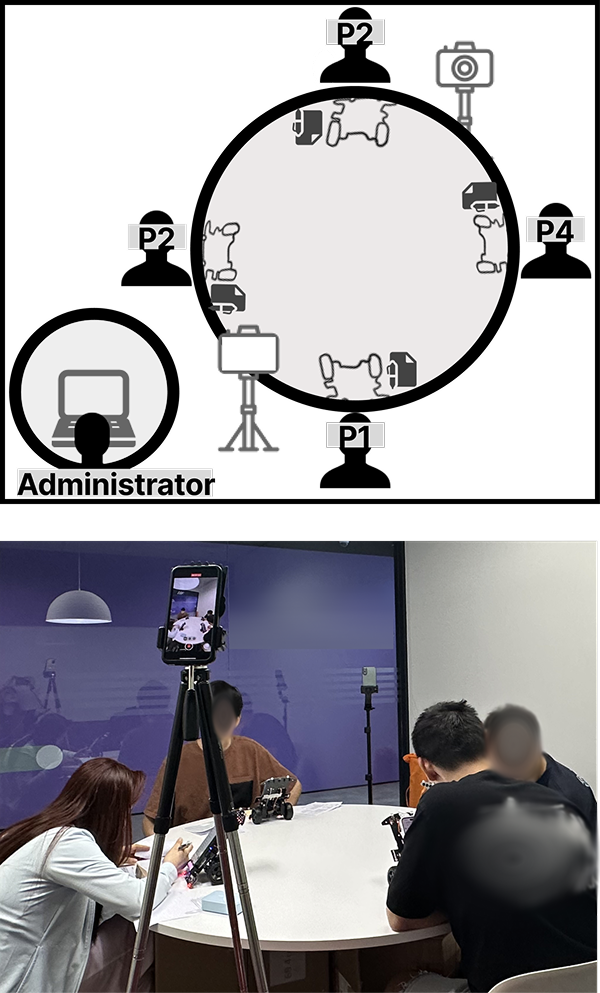}
    \caption{Setup of the experiment room. The left shows a bird's-eye view, and the right is the view from the administrator's view.}
    \label{fig:setup-exp}
    \end{subfigure}
    \hfill
    \begin{subfigure}[b]{0.5\textwidth}
        \includegraphics[width=\linewidth]{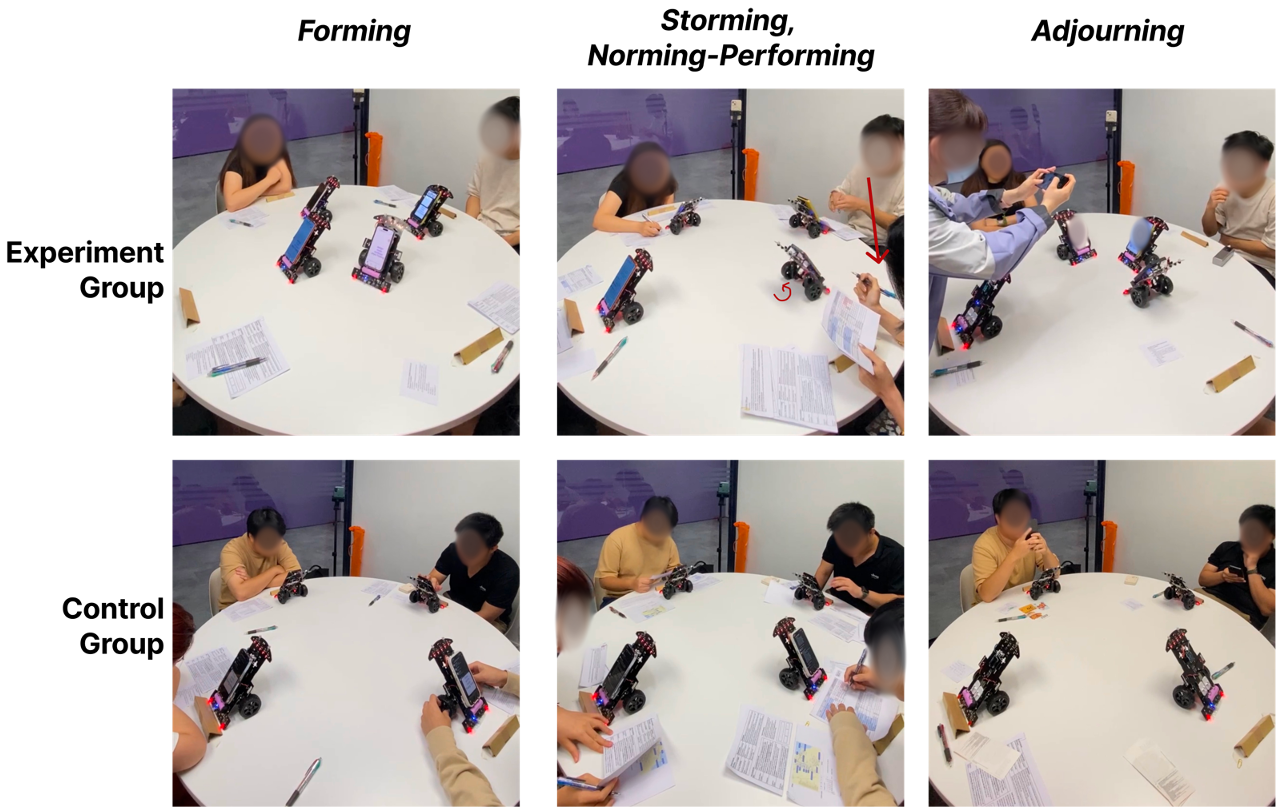}
        \caption{In the experiment groups with \textit{AnimaStand}'s facilitation, there were phone movements during all stages, while in the control groups, there were not. \\In the picture of \textit{Storming} and \textit{Norming-Performing} stage of the experiment group, the member sitting at the right top corner was asking the member sitting at the right bottom corner a question when their phone stepped out and rotated (marked in red line).}
        \label{fig:procedure}
    \end{subfigure}
    \hfill
    \begin{subfigure}[b]{0.28\textwidth}
        \includegraphics[width=0.9\linewidth]{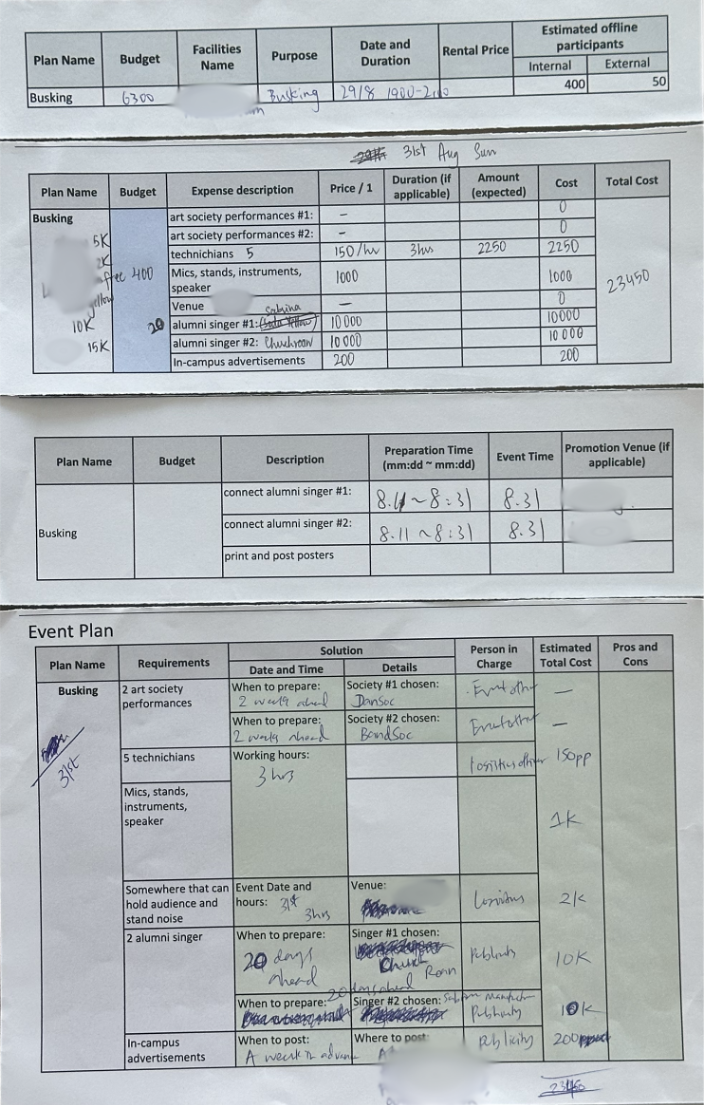}
        \caption{The example outcomes of the task, namely four completed individual task forms, each from one participant for one selected proposal.}
        \label{task_fig}
    \end{subfigure}
    \caption{The procedure and example outcomes of the experiment in five stages.}
    \Description{}
\end{figure}

The study took place in a room with a round table, where participants were evenly seated at four quadrants (Fig. \ref{fig:setup-exp}). The \textit{AnimaStand}s were positioned on the table facing the participants at a comfortable reading distance, with surrounding space for other printed materials. \textit{AnimaStand}s returned to their initial position after each movement. 
Participants used their own smartphones to log in to the Slack workspace (Sec.~\ref{slack}) and placed them on the \textit{AnimaStand} throughout the session. 
Three human administrators were seated in a corner: one hosted the session, one took notes, and the other operated the equipment (as described in Sec.~\ref{control-panel}). A laptop captured real‑time audio and ran the automatic problematic‑circumstance detection service. Two fixed cameras, placed at complementary angles, recorded all participants' behaviors.

\subsubsection{Experimental Conditions and Procedure}
After informed consent and voice registration for diarization, participants sat randomly around the table, placed their phones on \textit{AnimaStand}, and received a Slack introduction. Each posted their name and three self-descriptive keywords to generate an introduction image.

Both conditions received the same information and instructions. During \textit{Forming}, experimental groups' phones displayed introduction images while moving to face peers. All control phones displayed the same images at fixed intervals while remaining still. Administrators then introduced the task, roles, and bonuses. The 60-minute main session covered role and leader election followed by task completion. After role assignment, participants received private Slack channels and role-specific digital and printed materials. Experimental groups received \textit{AnimaStand} facilitation while control phones remained still. All groups received a 15-minute remaining-time alert.

At the conclusion, leaders presented their final proposals. During \textit{Adjourning}, all groups were instructed to exchange contacts: experimental phones moved outward to present social-media QR codes, while controls could use any method. Participants completed a questionnaire and a 10-minute interview (Table~\ref{tab:questionnaire_design}). Sessions lasted 120 minutes, with HKD 100 base compensation, a possible HKD 20 preference bonus, and an additional possible HKD 20 leadership bonus.

\subsection{Measures}\label{measures}
Drawing on Tuckman's group-development framework, we examined \textit{AnimaStand}'s effects on group collaboration (RQ2) in two dimensions: the \textbf{interpersonal dimension} (RQ2a), concerning how members acted toward and related to one another, and the \textbf{task-activity dimension} (RQ2b), concerning how they worked through the task. Each dimension included during-collaboration processes and post-collaboration outcomes measured through behavioral data, questionnaires, and interviews (details refer to appendix). We further examined how participants \textbf{perceived} (RQ3a) and \textbf{interpreted} (RQ3b) \textit{AnimaStand}.

\subsubsection{RQ2a: Interpersonal Dynamics and Relationships}
Following SYMLOG methodology \cite{bales1982symlog}, we analyzed verbal interaction quantitatively using speech diarization and nonverbal behavior qualitatively using session recordings. Because nonverbal behaviors are difficult to quantify or interpret for intent, and interaction in our discussion task was primarily verbal, conversational dynamics supported statistical evaluation while nonverbal observations provided contextual evidence. 

\paragraph{$\blacktriangleright$ Overview: Stages and Overall Verbal Participation}
For each stage, we measured duration and \textbf{speech coverage}: the proportion of time with at least one speaker, ranging from 0 (silence) to 1 (continuous speech).
Based on the overview, we identified \textit{Norming-Performing} as the longest and most interaction-intensive stage of the task session, which may provide rich conversational data for finer-grained analysis. We therefore examined not only the amount of verbal interaction, but also how participation and turn-taking were distributed across members throughout this stage.

\paragraph{$\blacktriangleright$ During Collaboration: Participation and Interdependence Balance}
Based on pilot observations, we divided \textit{Norming-Performing} into three substages: \textit{initialization}, in which members individually reviewed their role-specific materials without discussion; \textit{regular operation}, from the first discussion involving more than two members until the ``15 minutes remaining'' alert; and \textit{countdown operation}, from this alert until task completion. The last substage could vary when groups completed the task early. To account for different group paces without imposing a fixed clock-time axis, we used these substages as analytic units and calculated two group-level measures:
\begin{itemize}
    \item \textbf{Normalized Speech Distribution Entropy} (balanced participation), which measures the evenness of speaking time across members from 0 (one speaker or no observable speech) to 1 (equal speaking durations):
    \[
    H^{*}_{\text{speech}}=
    \begin{cases}
        \frac{-\sum_{i=1}^{N}p_i\log_2p_i}{\log_2N}, & \text{if }\sum_iT_i>0\\
        0, & \text{otherwise,}
    \end{cases}
    \quad p_i=\frac{T_i}{\sum_jT_j}.
    \]
    \item \textbf{Normalized Turn-taking Entropy} (balanced conversational interdependence), which measures the diversity and evenness of directed speaker transitions, from 0 (no observable turns or turns concentrated within one pair) to 1 (turns equally distributed across all possible directed speaker pairs):
    \[
    H^{*}_{\text{turn}}=
    \begin{cases}
        \frac{-\sum_a\sum_{b\neq a}q_{a\to b}\log_2q_{a\to b}}{\log_2[N(N-1)]}, & \text{if turns occur}\\
        0, & \text{otherwise,}
    \end{cases}
    \quad q_{a\to b}=\frac{C_{a\to b}}{\sum_i\sum_{j\neq i}C_{i\to j}}.
    \]
\end{itemize}
Here, $T_i$ is member $i$'s speaking time, $p_i$ is their proportion of total speaking time, $C_{a\to b}$ is the number of transitions from speaker $a$ to $b$, and $q_{a\to b}$ is that transition's proportion of all turns. 

Importantly, these measures characterize how evenly members participated and exchanged turns; they do not assume that greater balance necessarily indicates better collaboration. To contextualize the metrics, we examined behaviors induced by \textit{AnimaStand} according to facilitation type, actor, response latency, and behavior type, with supporting details (Sec.~\ref{data-analysis}).

\paragraph{$\blacktriangleright$ During Collaboration: Socializing Moments}
From recordings, we examined lively moments around facilitations designed for socializing (i.e., \textit{Icebreaking}, \textit{Conflict-Solving}, and \textit{Contact-Exchanging}), as well as other facilitations that unexpectedly elicited social responses. We recorded the facilitation type, members involved, and resulting behaviors. Interviews further captured participants' experiences of the collaboration climate, memorable moments, and impressions of other members.

\paragraph{$\blacktriangleright$ Post Collaboration: Relationship Assessments}
We assessed relationships at both dyadic and group levels. \textbf{Peer contribution evaluations}, obtained by distributing 100 points among all four members including oneself, captured recognition of session-specific contributions. 
\textbf{Oneness} then captured relationships beyond task contribution. Following prior work \cite{gachter2025measuring}, we calculated oneness by combining the \textit{IOS} and \textit{We} scales: 
\[
\text{Oneness}_{ij}=\frac{\text{IOS}_{ij}+\text{WeScale}_{ij}}{2},\quad
m_i=\min_{j\neq i}\text{Oneness}_{ij},\quad
\text{Oneness}_g=\frac{1}{N}\sum_{i=1}^{N}m_i.
\]
Here, $N=4$; $\text{Oneness}_{ij}\in[1,7]$ describes member $i$'s perceived \textbf{dyadic oneness} with member $j$; $m_i$ is member $i$'s lowest dyadic rating; and $\text{Oneness}_g$ is the group score. By aggregating each member's weakest perceived tie, \textbf{group oneness} captures the group's baseline, or ``minimum envelope,'' of oneness. This operationalization reflects the view that weaker interpersonal connections can constrain collective cohesion even when most dyadic relationships are strong.

\subsubsection{RQ2b: Task-Activity Progress and Outcomes}
We examined group task-operation progress during the core \textit{Norming-Performing} stage and assessed task performance after collaboration using the completed task materials and questionnaires.

\paragraph{$\blacktriangleright$ During Collaboration: Task Operation}
We characterized task progress through substage durations and transitions, alongside two content milestones including \textit{information alignment}, when members exchanged role-specific information, and \textit{proposal-completion discussion}, when they elaborated initial proposals one at a time (Sec.~\ref{task-design}). These measures and behavioral observations informed our analysis of task-operation strategies.

\paragraph{$\blacktriangleright$ Post Collaboration: Task Performance}
We evaluated task performance objectively and subjectively. Objective \textbf{task completion rate} was the proportion of filled blanks in the submitted plan form (Fig.~\ref{task_fig}). Participants also rated their \textbf{confidence} in and \textbf{satisfaction} with the outcome on 5-point Likert scales.

\subsubsection{RQ3a: Perception of \textit{AnimaStand}}
Using 5-point Likert items adapted from prior work on social robots and interfaces, we measured perceived \textbf{interpretability}~\cite{10.1145/3613905.3651027}, \textbf{influence}~\cite{10.1145/3613905.3651027}, \textbf{distraction}~\cite{bell2023evidence}, \textbf{helpfulness}~\cite{bell2023evidence}, and \textbf{likeability}~\cite{bartneck2009measurement}. Interviews probed the reasons for these ratings and participants' suggestions and expectations for future development.

\subsubsection{RQ3b: Interpretation of Facilitation Movements}
To examine possible links between interpretation and behavioral effects, we identified repetitive or typical reactions to facilitation in session recordings. At the end of the interview, participants first explained their reactions and then interpreted the corresponding movements without being told the intended meanings.

\subsection{Data Collection and Analysis}\label{data-analysis}
We collected 14 session recordings with speech diarization (Sec.~\ref{4.3.1}), questionnaires, and transcribed semi-structured interviews. 

\textbf{\textit{Quantitative analysis.}}
Using diarization data, we calculated speech coverage, then averaged sliding-window speech and turn-taking entropies within each \textit{Norming--Performing} substage. Questionnaire ratings were analyzed directly, except IOS and We scores, which were combined into oneness. 
Significance was assessed using Welch's t-test for normally distributed data and the Mann–Whitney U test otherwise. All statistical tests were two-sided. We report Cohen's $d$ for Welch's t-tests and rank-biserial $r$ for Mann--Whitney U tests. Given the exploratory design and limited number of groups, p-values were not adjusted for multiple comparisons and were interpreted alongside effect sizes rather than as confirmatory evidence.

\textbf{\textit{Qualitative analysis.}}
Two authors conducted mixed-methods content analysis of videos \cite{creswell2017designing, hsieh2005three}, coding stages and substages with calculation of their durations, facilitation-triggered behaviors, and task milestones (Sec.~\ref{measures}). 
Three authors then coded interview transcripts and organized findings by research question. It is worth noting that only experimental participants' responses informed the perception and interpretation analyses. Control participants answered the same items for procedural consistency; these items appeared last to avoid influencing other responses.
For all coding procedure, the authors iteratively compared annotations, refined category definitions, and resolved disagreements through discussion.

%% file: 6-results-new.tex
\section{Results: Effects and Perception of Animated Phone Facilitation}
In this section, we present the effects of \textit{AnimaStand}'s facilitation in terms of during-collaboration dynamics and post-collaboration outcomes (\textbf{RQ2}). For each analytic dimension, we first report quantitative patterns and then provide behavioral analyses to contextualize how these patterns emerged. Given the limited number of groups, statistical comparisons are treated as exploratory evidence rather than confirmatory tests.
Then, we present findings regarding participants' perceptions and interpretations of \textit{AnimaStand} (\textbf{RQ3}). 

\subsection{Interpersonal Dynamics and Relationship (RQ2a)}
We first examined how \textit{AnimaStand} affects interpersonal collaboration, including during-collaboration interpersonal dynamics and post-collaboration perceived relationship.
\subsubsection{Overview: Overall Level of Participation Across Stages}
\begin{figure}[h]
    \centering
    \includegraphics[width=0.65\linewidth]{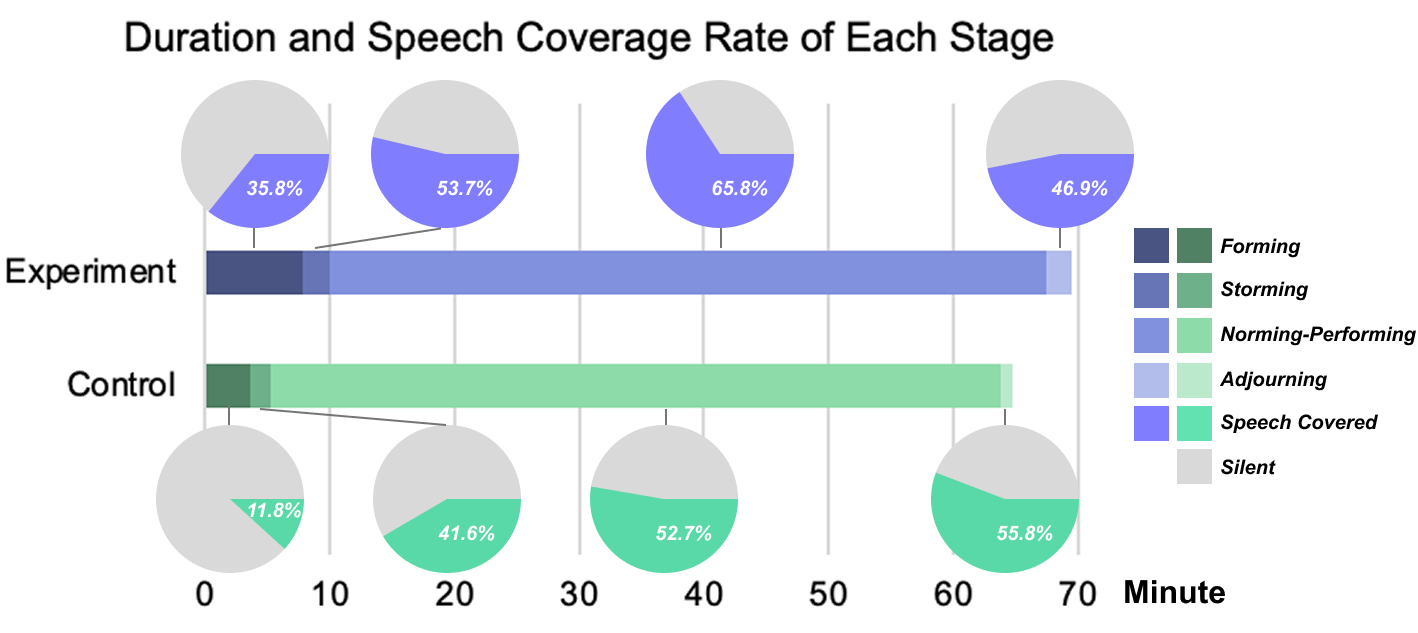}
    \caption{Duration and speech coverage rate of each stage. The bar chart demonstrates the duration, while the pie charts linked out above and below present the speech coverage rate, respectively.}
    \Description{}
    \label{fig:duration_active}
\end{figure}

To provide a comprehensive view, we first compared the speech coverage rate of each Tuckman stage between conditions, serving as a basis for further analysis (Fig.~\ref{fig:duration_active}). The \textit{Norming} and \textit{Performing} stages are combined, consistent with the design workshop. Participants from experimental groups were generally participating more actively at each stage than those from control groups, as measured by speech coverage rate. Experimental groups showed a significantly higher speech coverage rate in the \textit{Forming} stage ($ 0.36 \pm 0.22$ vs. $ 0.12 \pm 0.13$, $p = .026$, $ d = 1.36$) and marginally higher in the \textit{Norming-Performing} stage than the control condition ($ 0.66 \pm 0.15$ vs. $0.53 \pm 0.13$, $p = .099, d = 0.96$). 

In \textit{Forming}, this higher speech coverage was likely related to the \textit{Icebreaking} and \textit{Speech-Control} facilitations. For \textit{Icebreaking}, phones gathered and rotated together at the center, then facing the audience one by one as a digital name tag. Such movements caused six of seven experimental groups' members to begin introducing themselves after a brief pause to jointly interpret this cue. In contrast, although the control group's phones displayed the same personal information in Slack's public channel, the absence of movement led some participants to only type in Slack instead of speaking, while others remained silent, ``\textit{If not the Slack where we type to chat, I actually do not know how to start up a conversation}''(C3P2). The \textit{Speech-Control} facilitation prompted members whose introductions lasted under 15 seconds with back‑and‑forth motion, futher encouraging richer self‑presentations. For instance, E2P4 urged E2P3 to ``\textit{say more,}'' after witnessing the facilitation, leading them to expand on their keywords. 

\subsubsection{During-Collaboration: Balance of Participation and Interdependence}\label{during-collaboration-interpersonal}

\paragraph{\textbf{\textit{AnimaStand} Balanced Participation and Interdependence during the \textit{Norming–Performing} Stage}}
We compared speech-distribution and turn-taking entropies across the \textit{initialization}, \textit{regular operation}, and \textit{countdown operation} substages of \textit{Norming--Performing} (Sec.~\ref{measures}).

\begin{figure}[htbp]
  \centering
  %\begin{subfigure}[t]{0.32\textwidth}
    %\includegraphics[width=\linewidth]{fig/Speech-Coverage-Rate-barplot-with-table-1119.png}
    %\caption{Speech coverage rate.}
    %\label{fig:sub-active}
  %\end{subfigure}
  %\hfill
  \begin{subfigure}[t]{0.4\textwidth}
    \includegraphics[width=\linewidth]{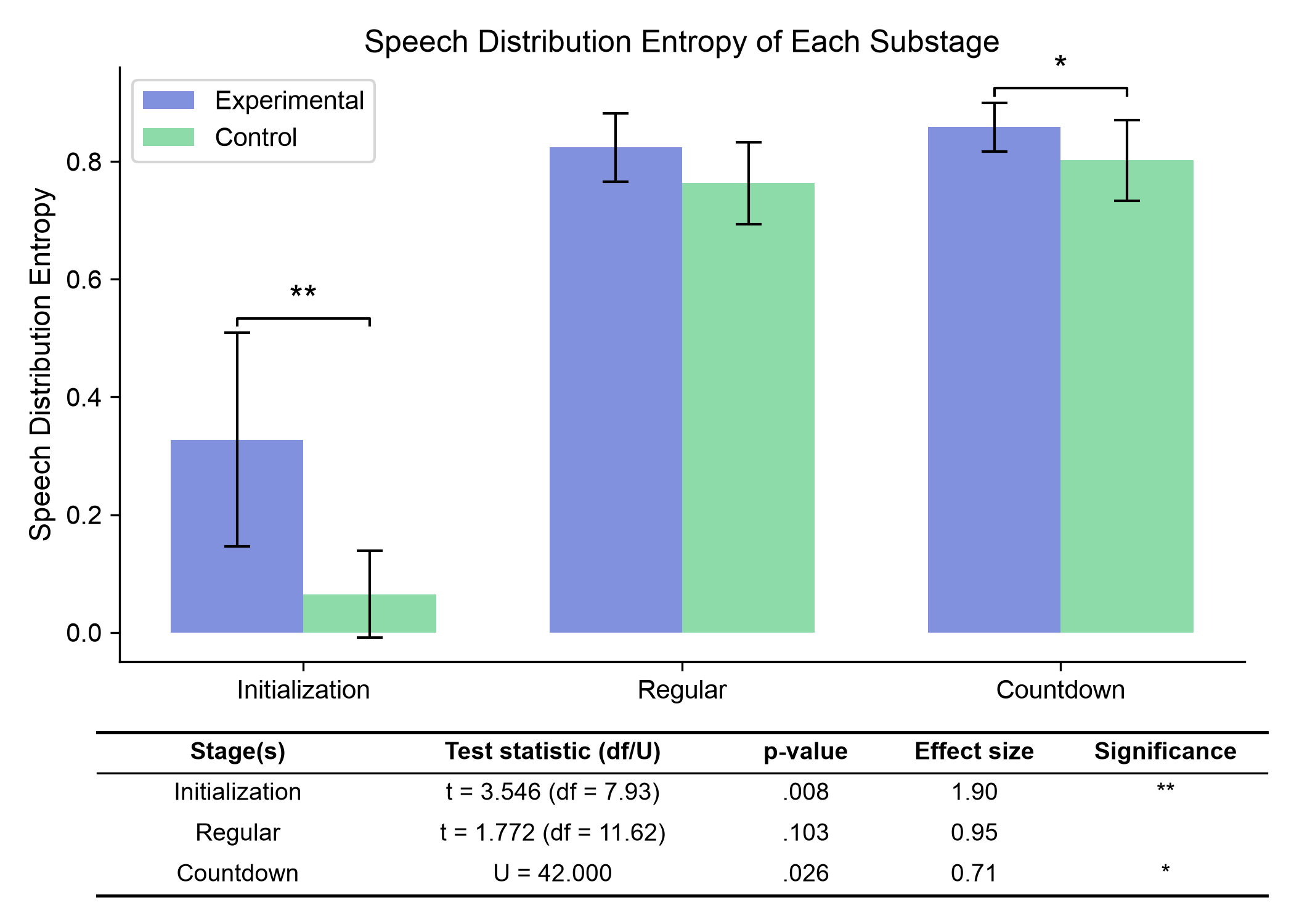}
    \caption{Speech distribution entropy.}
    \label{fig:sub-speech}
  \end{subfigure}
  \begin{subfigure}[t]{0.4\textwidth}
    \includegraphics[width=\linewidth]{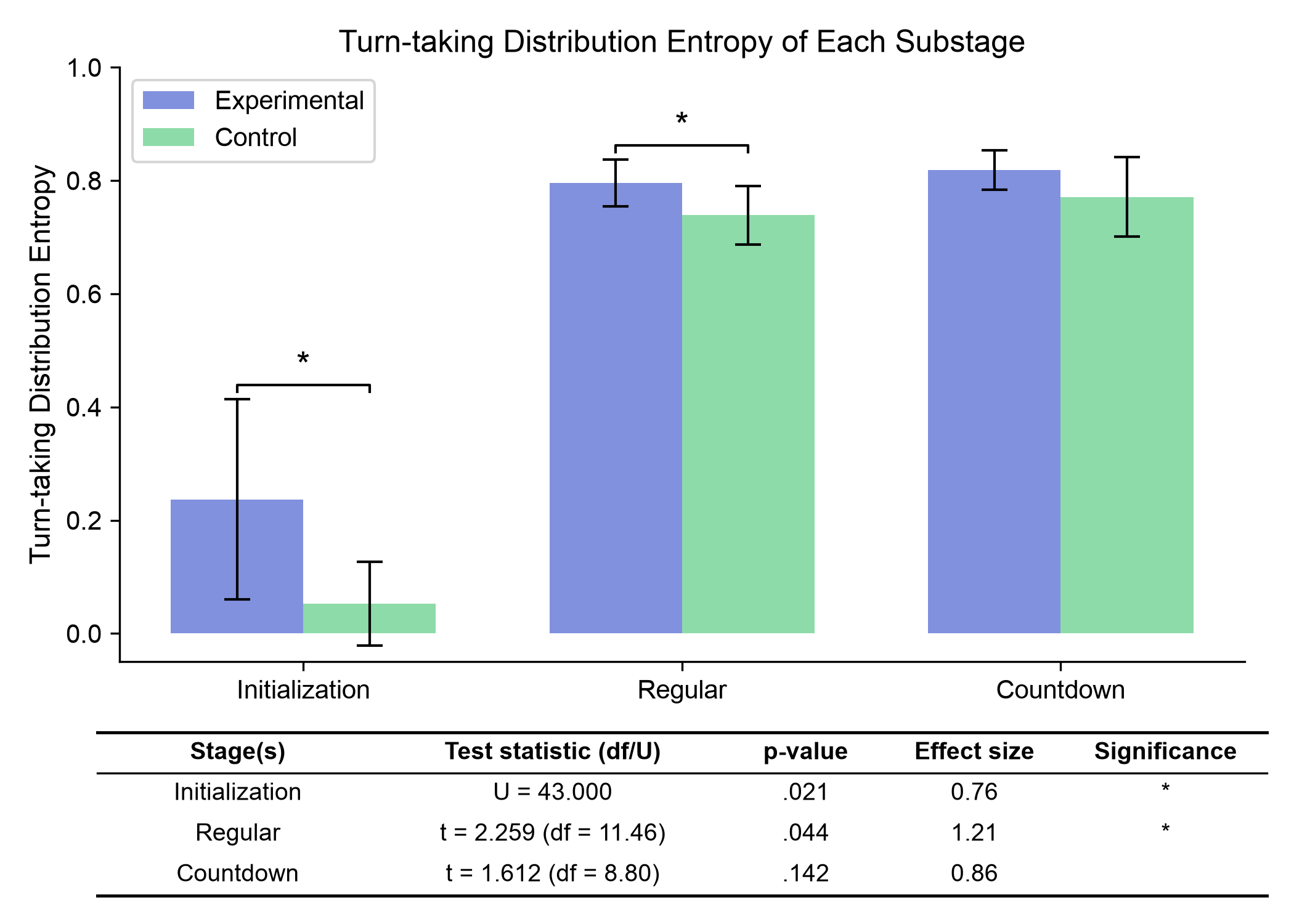}
    \caption{Turn-taking distribution entropy.}
    \label{fig:sub-turn}
  \end{subfigure}
  \captionsetup{skip=0.5ex}
  \caption{The level of balance of participation and interdependence, indicated by speech distribution entropy, and turn-taking distribution entropy. 
  Error bars depict standard deviations.
  The Welch's t-test (normal distribution) or Mann-Whitney U-test (otherwise) was employed for between-group comparisons.
  Significance values are reported for 0.05 < p < 0.1 ( $\hat{}$ ), p < .05 (*), p < .01 (**), p < .001 (***), abbreviated by the number of stars. We calculated and presented Cohen's d (t-test) or rank-biserial r (U-test) as an indicator of effect size for significant comparisons.}
  \label{fig:participation-interdependence}
\end{figure}

As in Fig.~\ref{fig:participation-interdependence}, in the \textit{initialization} substage, experimental groups exhibited significantly higher values in all metrics than controls, a difference likely attributable to the control groups' prolonged initialization periods with extended silences, which suppressed speech coverage and entropy measures. 
%\lzq{Some t, df, and p values in the plots do not match. For example, t=2.276 with df=12 gives a two-sided p of about .042, not .022. Check the test direction and df against the analysis output.}
During \textit{regular operation}, experimental groups scored higher on all metrics, with turn‑taking entropy reaching statistical significance ($p = .044, d = 1.21$).
In the \textit{countdown operation} substage, experimental groups again outperformed controls across all metrics, with speech distribution entropy showing significance ($ p = .026, r = 0.71$).
These differences may be related to \textit{Participation-Balance} facilitations, which are the most frequently triggered ones in experimental groups during \textit{Norming-Performing} when they encountered participation imbalances like uneven speaker dominance. We next examine how these cues elicited members' behavioral engagement and contributed to more balanced participation and turn-taking.

\begin{figure}[ht]
    \centering
    \includegraphics[width=0.7\linewidth]{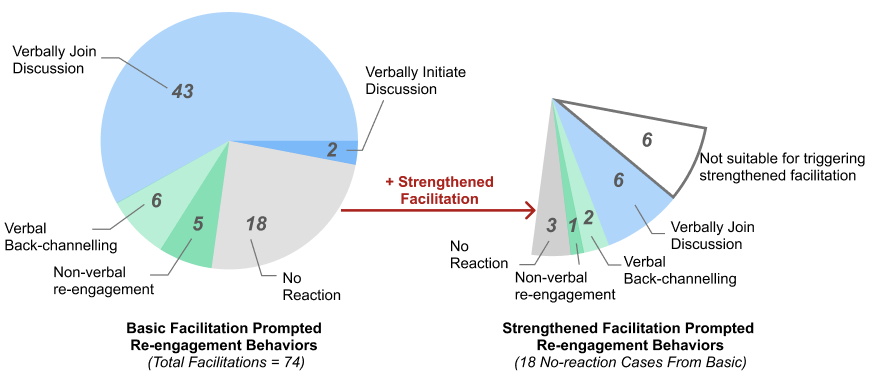}
    \caption{Prompted group work–related behavior types and their distribution triggered by basic and strengthened Participation‑Balance facilitations. The left shows behaviors triggered by basic facilitation; the right shows those triggered by strengthened ones (18 cases where basic facilitation previously elicited no reaction).}
    \label{fig:prompted-behavior-distribution}
\end{figure}

\paragraph{\textbf{\textit{AnimaStand}'s \textit{Participation-Balance} Cues Primarily Promoted Verbal Re-engagements, While Also Eliciting Non-Verbal Responses}}\label{reengagement-types}
First, we coded inactive members' responses to the facilitations into re-engagement types (Fig. \ref{fig:prompted-behavior-distribution}).
In our coding, the non-verbal re-engagement only counted those reactions that showed observable attention shifts to the group discussion, such as leaning in to follow ongoing discussions (E3P1).
In the experimental condition, basic \textit{Participation–Balance} facilitation (i.e., less active members' phones to step out and rotate) was triggered 74 times, re-engaging inactive members in 56 cases to rejoin (51 verbally and 5 non-verbally). 
In the remaining 18 cases, six were unsuitable for strengthened facilitation (e.g., phones physically obstructed), while the other 12 triggered the strengthened facilitation (active member's phone moving toward the least active and pulling it toward the center), with nine resulting in re-engagement (six verbal, two verbal back-channeling, and one non-verbal).
The re-engagement types and distributions prompted by basic and strengthened facilitation were largely similar.

\paragraph{\textbf{\textit{AnimaStand}'s Inviting Movements Encouraged Salient Verbal Engagement from Both Inactive and Active Members, Balancing Participation and Interdependence}}
To reveal the mechanism by which these interventions influenced the members and ultimately balanced the groups' dynamics (as in Fig.~\ref{fig:participation-interdependence}), we first focus on the dominant verbal re-engagement (45/56 prompted by basic facilitation, 6/9 by strengthened facilitation), excluding brief verbal back-channeling.

% interesting thing 1: who actually invited them.
Firstly, we discovered that \textbf{the moving phones not only prompted their inactive owners, but also other active members}. Inactive members responded directly as the first speaker in 36 instances (31 basic, five strengthened). Alternatively, in 13 other instances (12 basic, one strengthened), an active member noticed the facilitation and explicitly invited the inactive member to join first, then the inactive ones followed.
For example, after E7P3's phone rotated, E7P4 invited them within 13 seconds (including 8 seconds of phone movement), prompting immediate engagement. Similarly, E6P1 always (5 out of 9 facilitations) initiated talk with the inactive upon noticing such phone behaviors. Such inviting behavior between members created more opportunities for them to release prosocial signs to others, which may have led to enhanced relationships \cite{caputi2012longitudinal}.

\begin{figure}[htbp]
    \centering
    \begin{subfigure}[t]{0.31\textwidth}
        \includegraphics[width=\linewidth]{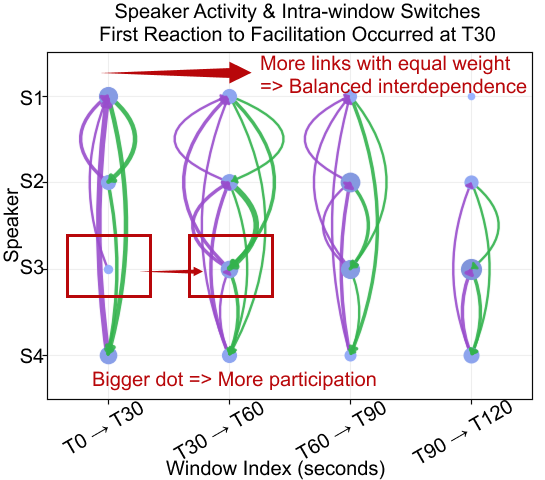}
        \caption{Type 1.1: Active Verbal Re-engagement (\textit{\textbf{fair speech example}}, S3 is target, effective re-engagement occurred immediately)}
        \label{fig:active-verbal-rejoin}
    \end{subfigure}
    \hfill
    \begin{subfigure}[t]{0.31\textwidth}
        \includegraphics[width=\linewidth]{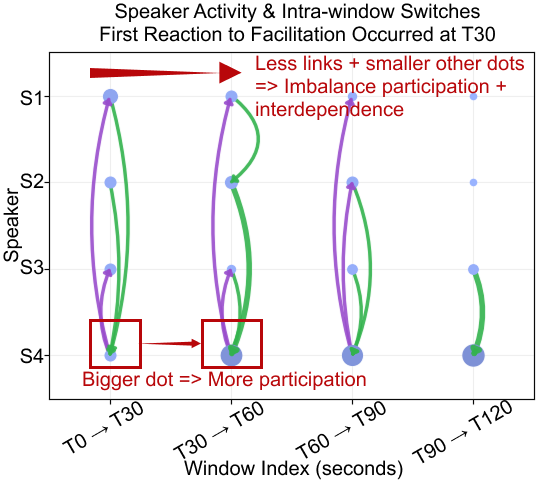}
        \caption{Type 1.2: Active Verbal Re-engagement (\textit{\textbf{reactive over-speech example}}, S4 is target, re-engagement occurred immediately; however, the target engaged too much, causing another imbalance)}
        \label{fig:reactive-over-rejoin}
    \end{subfigure}
    \hfill
    \begin{subfigure}[t]{0.31\textwidth}
        \includegraphics[width=\linewidth]{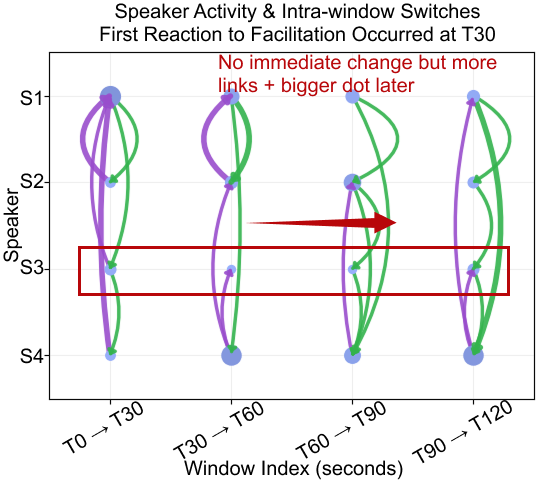}
        \caption{Type 2: Non-active/Non-verbal re-engagement example (S3 is target, effective re-engagement did not occur immediately; instead, the target re-engaged after a latency)}
        \label{fig:non-verbal-rejoin}
    \end{subfigure}
    \captionsetup{skip=0.5ex}
    \caption{Speaker activity patterns before and after a Participation‑Balance facilitation. Lines on the left indicate turn‑taking from members with smaller to larger IDs (e.g., Member 1 followed by Member 4), while lines on the right indicate the reverse. Line thickness represents the frequency of turn‑taking within the time window, and dot size represents total speaking duration; thicker lines indicate higher frequency, and larger dots indicate longer durations.}
    \label{fig:rejoin-pattern}
\end{figure}

% interesting 2 结合dynamic变化图解释
Moreover, regardless of who initiated the speech after the facilitation, we discovered that \textbf{when the inactive members rejoined the discussion, they re-balanced both the participation and interdependence} (Fig.~\ref{fig:active-verbal-rejoin}).
When re-engaged, they often tried to offer adequate input, thereby increasing their speaking time (the dot size representing speech time grows larger after re-engagement).
Since inactive members also had withheld unique knowledge during their prolonged silence, others often expected more information exchange with them once they were ``back-online''.
This often triggered turn-taking between the inactive and other members (more links appear after re-engagement), increasing and diversifying directed speaker transitions.
%Their previous silence also made them focal points for information retrieval, as others had not sought input from them for some time. 
However, in some cases, participation balance did not improve within the next few minutes. We observed that some previously inactive members, once prompted, tended to engage in reactive over‐speaking—continuously talking for an extended period after a long silence, as shown in Fig.~\ref{fig:reactive-over-rejoin}. This ironically made them the most dominant speaker within the surrounding windows and thus reduced the local participation balance.

\paragraph{\textbf{\textit{AnimaStand} Evoked Light Verbal and Non-Verbal Re-engagement, Preparing Inactive Members for Future Discussion}}\label{noise}
The 14 other occasions where the inactive members responded to the engagement invitation only through verbal backchanneling (e.g., ``\textit{ok, I got you}'' by E4P2) or non-verbal attempts (e.g., attention shift to the current speaker by E2P1) are also remarkable.
Although their responses did not directly balance participation and interdependence between members, \textbf{these light verbal and non-verbal re-engagements sometimes successfully redirected inactive members' attention to the collaboration and set them ``on call''}, increasing readiness for subsequent participation and potentially supporting smoother turn-taking later (as in Fig.~\ref{fig:non-verbal-rejoin}).
However, there are also other cases where users were puzzled (3 occasions), sensing they ``\textit{should do something now}'' but unsure how (E5P1), and therefore reacted slightly to avoid embarrassment. While these responses increased observable activity, they did not substantively advance the discussion but rather appeared as interactional ``noise''.
% Such reactions were rooted in these inactive participants' interpretation of the facilitation and their perceived situational constraints.
% Some understood the prompt (7 cases) but could not immediately join actively due to ongoing conversation flow (E3P1) or lack of relevant input (E4P2). Some misinterpreted the signal (4 cases), as E1P1 read phone rotation as a cue to ``\textit{provide more information}'' and began searching materials for future usage instead of speaking. Others were puzzled (3 occasions), sensing they ``\textit{should do something now}'' but unsure how (E5P1).

\paragraph{\textbf{\textit{AnimaStand}'s Facilitation Effects Persisted When Intervention Frequency Dropped During the Countdown Operation}}\label{persist-effect}
Upon receiving the ``last 15 minutes'' warning, all groups, experimental or control, expressed surprise at the limited time remaining and accelerated their task progress. 
In reaction to the reminder of the time limit, in experimental conditions, speech distribution entropy rose significantly (paired-test $p = .016, r = 1.00$); while the occurrence of phone facilitation in experimental groups dropped from an average of $4.81 \pm 0.95$ times per 15 minutes to $2.29 \pm 1.50$ times per 15 minutes.
%In reaction to the reminder of the time limit, in experiment conditions, speech coverage rate and speech distribution entropy rose significantly (paired-test $p = .01, d = .39$, $p = .03, d = 1.00$); while the occurrence of phone facilitation in experiment groups dropped from an average of $4.81 \pm 0.95$ times per 15 minutes to $2.29 \pm 1.50$ times per 15 minutes.
%In contrast, control groups showed no significant improvement in all metrics, even a drop in speech coverage rate (Fig.~\ref{fig:participation-interdependence}).
In contrast, control groups showed no significant change in all metrics (Fig.~\ref{fig:participation-interdependence}).
This pattern suggests that the tendency toward more evenly distributed participation and turn-taking associated with animated-phone interventions during \textit{regular operation} may have persisted into \textit{countdown operation} despite fewer interventions, whereas no comparable carry-over pattern was observed in control groups.
It may be because the participants in the experimental condition already formed the impression that they should ``\textit{at least say something, as we have been reminded throughout}''(E5P1). 
On the contrary, members in the control groups did not develop such an awareness to be more active, and thus ``\textit{As the task proceeded to the last minutes, seeing the growing pile of todos, I felt desperate and did not know what to contribute}''(C3P3).

\subsubsection{During-Collaboration: Socializing Moments}
\paragraph{\textbf{\textit{AnimaStand}'s Synchronous Movements Added Lively Vibe and Emphasized Cohesion}}
Although most of \textit{AnimaStand} facilitations are not designed to elevate the atmosphere, many of them also stimulated socializing responses. 
In particular, this often happened in facilitations where phones moved in synchronized group formations. 
For instance, during \textit{Forming}, participants' first conversations are often concerned with jointly interpreting the intention and metaphor behind the \textit{Icebreaking} facilitation. Similarly, in \textit{Norming-Performing}, as phones moved synchronously in the \textit{Silence‑Breaking} facilitation, participants exhibited a series of notable immediate reactions, including describing the movement, speculating about its intention or purpose, and expressing amusement through laughter or chuckles. These actions helped alleviate the awkward atmosphere created by the prolonged silence. Some even cited this as the most pleasant moment of the session (E1P1\&4, E2P2, E3P1, E4P1), highlighting its value as both a progress and social trigger.

\paragraph{\textbf{\textit{AnimaStand}'s Public Intervention Invited Joint Effort to Solve Relational Problems in Some Cases.}}\label{conflict}
Only two conflicts occurred, both in experimental groups. After E2P3 and E2P4 debated a venue for over ten exchanges, \textit{Conflict-Solving} movements redirected their attention, and they deferred to E2P3's venue-related role. In the other case, E6P1 interpreted the cue during E6P3 and E6P4's disagreement and proposed a vote.
These cases suggest that \textit{Conflict-Solving} facilitation, like \textit{Participation-Balance} facilitation, cues both the phones' owners and others to act. With publicly visible phone movements enabling all peers to intervene and share responsibility for group regulation, these facilitations cultivate the oneness within the group.
Comparatively, private intervention -- the \textit{Connection-Tickle} -- designed to address the lack of interdependence between two members, was difficult to conceive under the tight time constraints and was applied only twice by the same participant (E1P3): once during the \textit{Norming–Performing} stage and once in the \textit{Adjourning}, purely for amusement. 
Although private intervention showed limited practical utility in fast-paced, in-person discussions, it may still ease tension and foster a positive group atmosphere in a lighthearted way.

\paragraph{\textbf{In \textit{Adjourning}, \textit{AnimaStand} Facilitated Non-Verbal Contact Exchange, While Also Inducing Possible Over-Reliance}}\label{adjourning-bhv}
During \textit{Adjourning}, \textit{Farewell} facilitation emphasized non-verbal socializing by guiding participants to exchange contact information through phone movements and on-screen QR codes. Six of seven experimental groups successfully exchanged contact information, often accompanied by laughter and small talk, and one remaining group created a face‑to‑face chat group (using a shared 4-digit code to form a nearby group). In contrast, although control groups received the same instructions, only two eventually exchanged contact information, both via face‑to‑face group creation. This suggests that animated phone cues can support lightweight connection-building beyond verbal discussion. However, the less speech in experimental groups during \textit{Adjourning} may also indicate cue-waiting or over-reliance on facilitation, whereas less-guided control groups sometimes interacted more freely.

\subsubsection{Post-Collaboration: Relationship Assessments}
\begin{figure}[ht]
    \centering
    \begin{subfigure}[t]{0.29\textwidth}
        \centering
        \includegraphics[width=\linewidth]{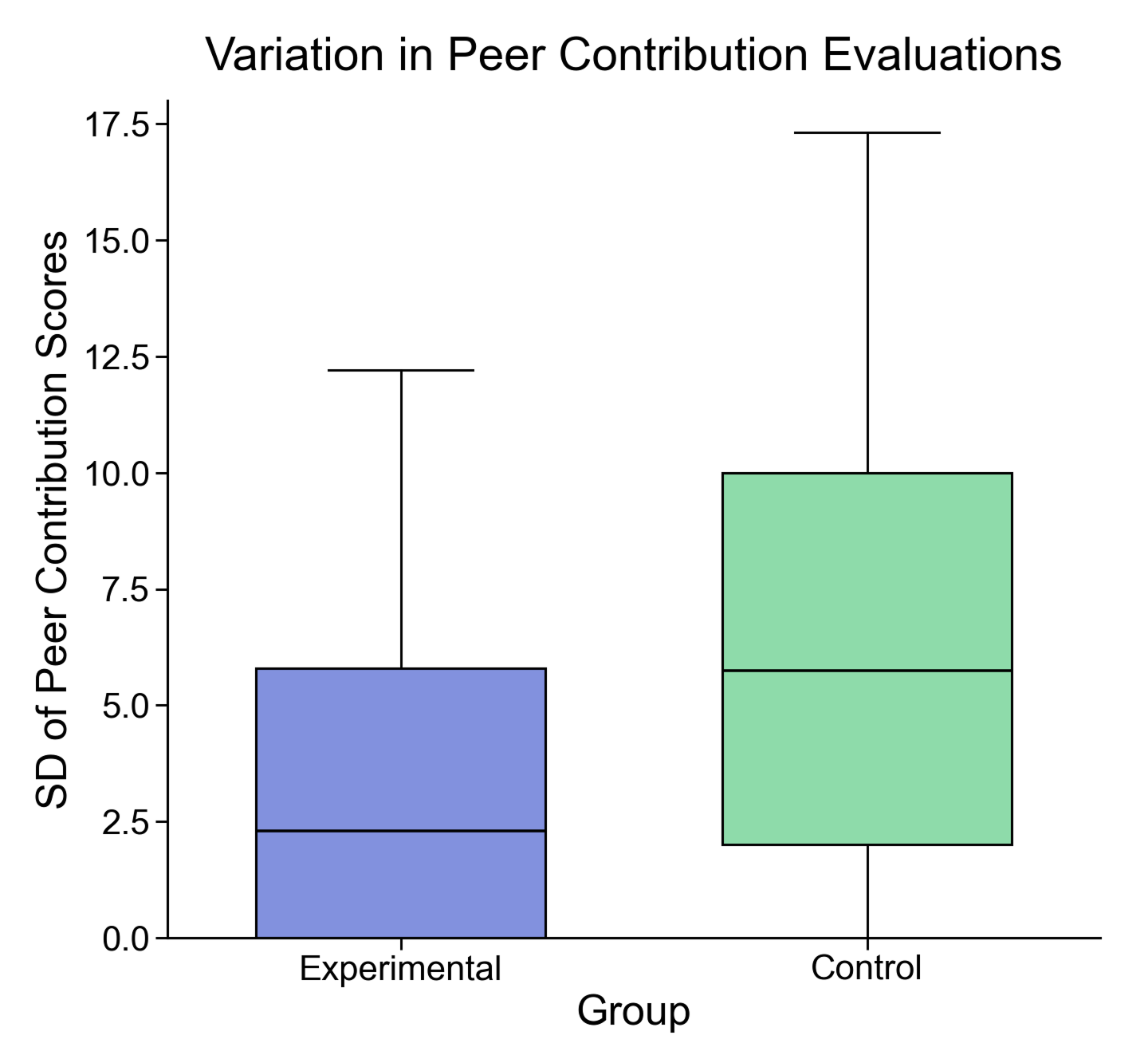}
        \captionsetup{skip=0.5ex}
        \caption{SD of peer evaluation (distributing 100 points among all four members, including themselves)}
        \label{fig:peer-eval}
    \end{subfigure}
    \hspace{0.04\linewidth}
    \begin{subfigure}[t]{0.33\textwidth}
        \centering
        \includegraphics[width=\linewidth]{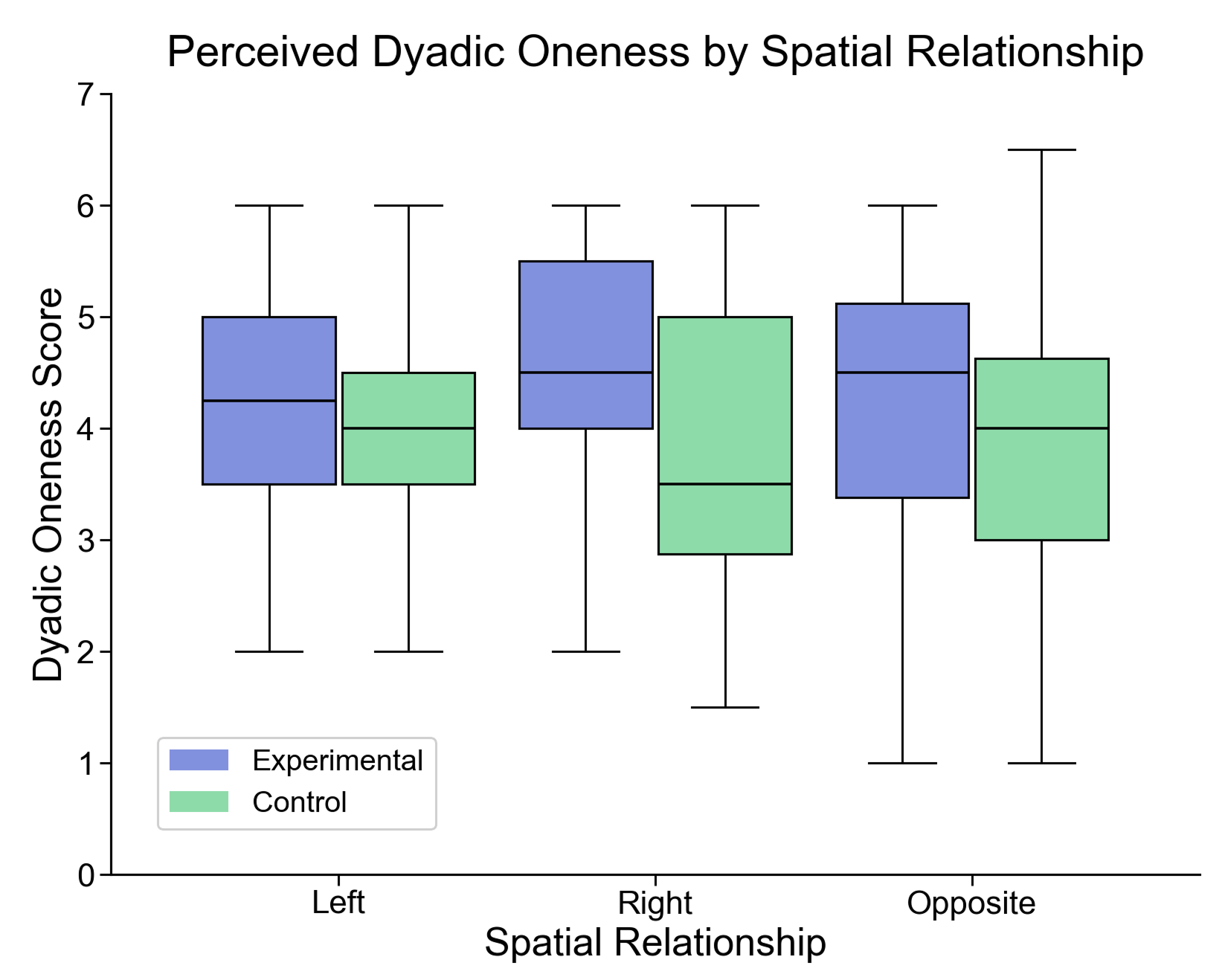}
        \caption{Individual Oneness with other members seated at different spatial positions around the round table}.
        \label{fig:oneness-indiv}
    \end{subfigure}
    \hspace{0.04\linewidth}
    \begin{subfigure}[t]{0.27\textwidth}
        \centering
        \includegraphics[width=\linewidth]{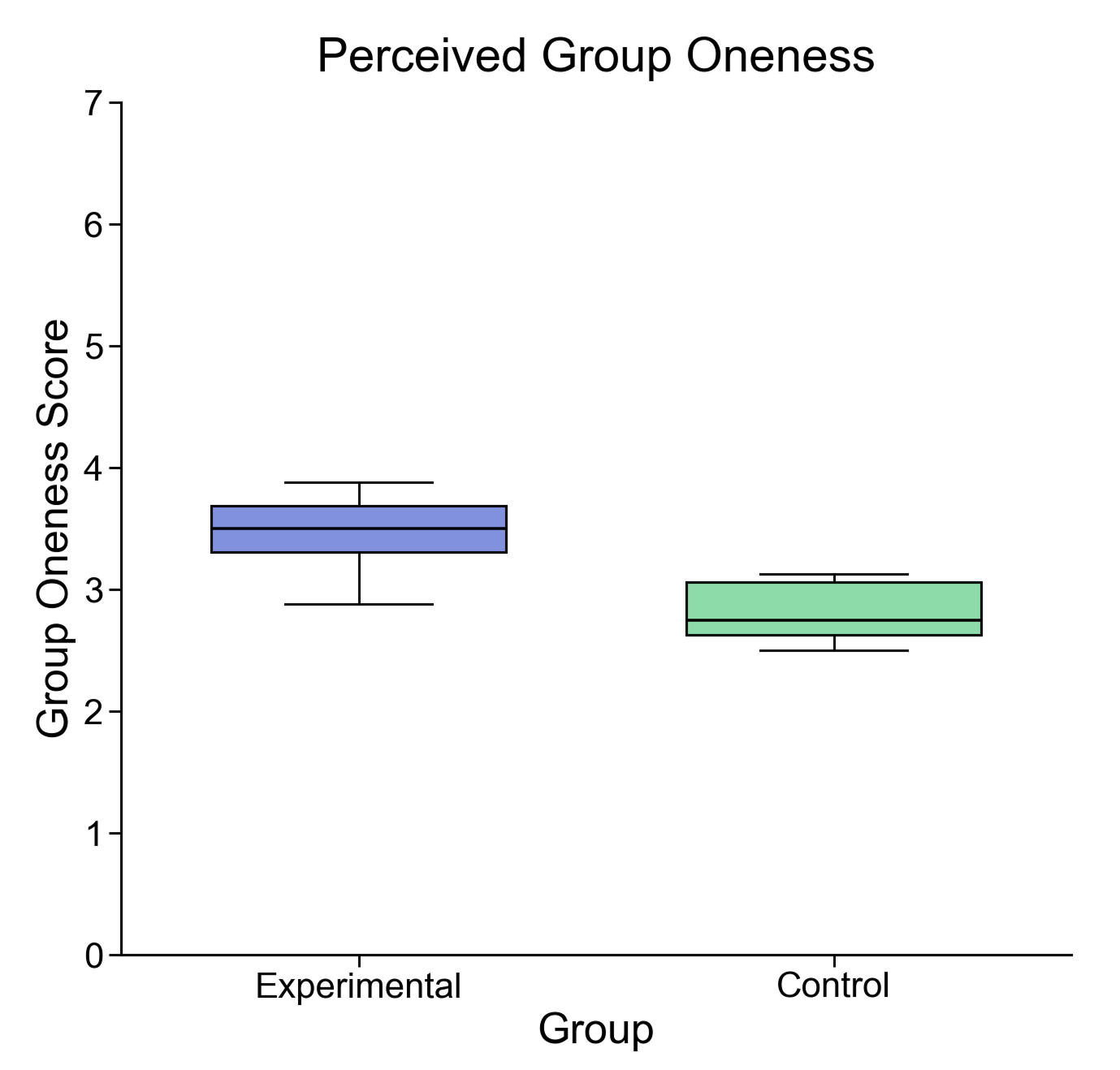}
        \caption{Group Oneness, calculated from the lowest individual rating within a group.}
        \label{fig:oneness-grp}
    \end{subfigure}
    \captionsetup{skip=0.5ex}
    \caption{Peer evaluation, perceived dyadic oneness, and group oneness. Error bars depict boxplot whiskers.}
    \Description{}
    \label{fig:oneness}
\end{figure}

\paragraph{\textbf{\textit{AnimaStand} Promoted More Even Peer-Contribution Recognition and Descriptively Higher Group Oneness}}
We first examined peer-level relationship assessments through peer contribution evaluations and dyadic oneness ratings. For peer contribution evaluations, participants from the experimental condition generally distributed the 100 points more evenly among members, with significantly lower standard deviation ($p = .026, r = .71$, considering their grouping).  
For dyadic oneness, ratings toward members seated at different spatial positions around the round table showed no significant difference (Fig.~\ref{fig:oneness-indiv}).

We then examined group-level relationship assessment using the aggregated group oneness score (Fig.~\ref{fig:oneness-grp}). The group oneness score was descriptively higher in experimental groups, although marginally ($p = .054, r = .63$), indicating slightly greater collective oneness. Because group oneness was calculated from each member's lowest dyadic oneness rating, this slightly higher score suggests that \textit{AnimaStand} might have improved the weakest perceived ties within the group, thereby raising the baseline level of group cohesion.

\subsection{Task-Activity Progress and Outcomes (RQ2b)}
We then examined how \textit{AnimaStand} shaped task activity, particularly focusing on task-operation progress during \textit{Norming-Performing} and post-collaboration task outcome.

\subsubsection{During-Collaboration: Task Operation Progress and Strategy}
\begin{figure}[ht]
    \centering
    \begin{subfigure}{0.4\textwidth}
    \includegraphics[width = \linewidth]{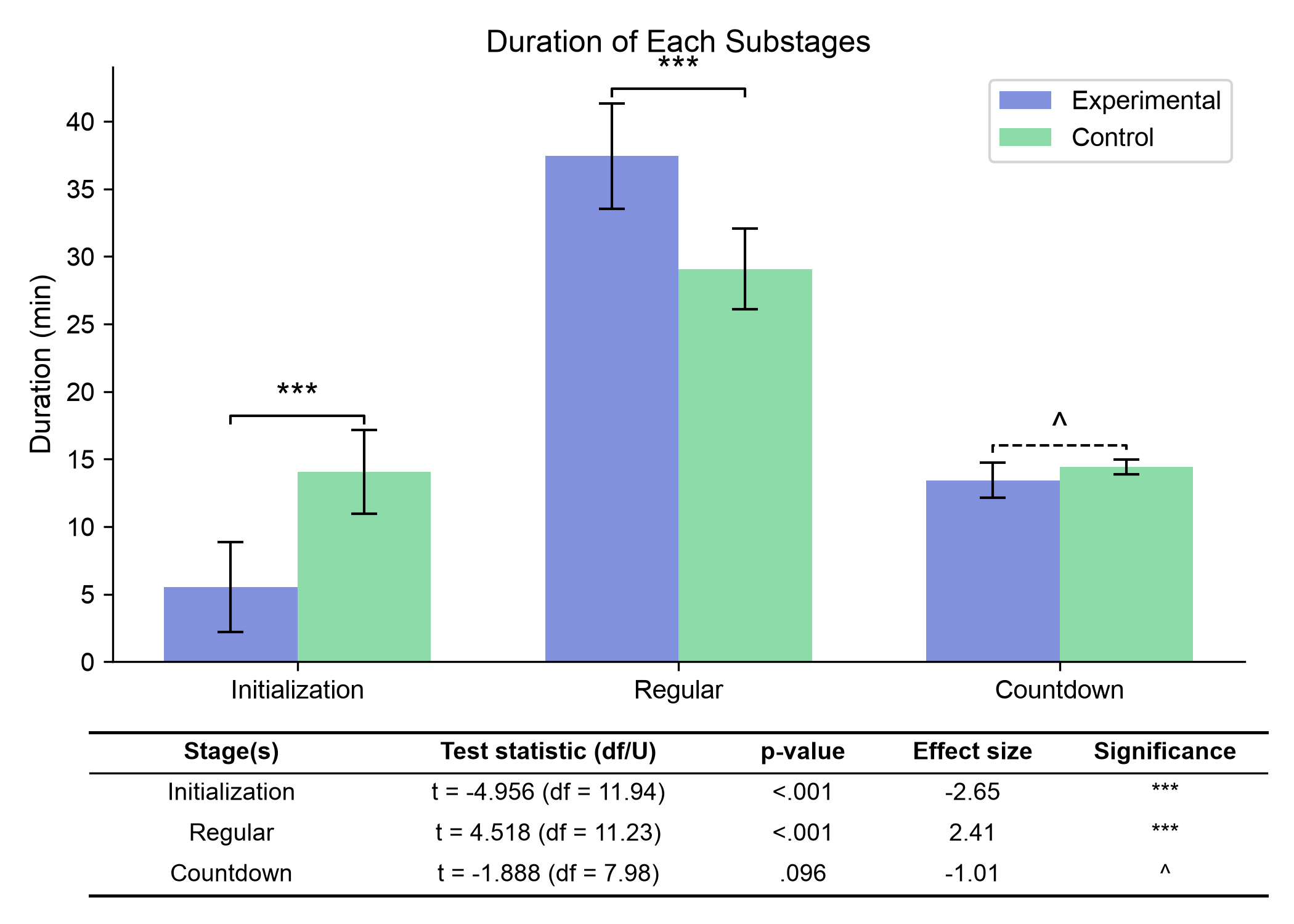}
    \caption{The duration of three substages in experimental and control groups.}
    \Description{}
    \label{fig:sub-duration}
    \end{subfigure}
    \hfill
    \begin{subfigure}[b]{0.50\textwidth}
        \includegraphics[width=\linewidth]{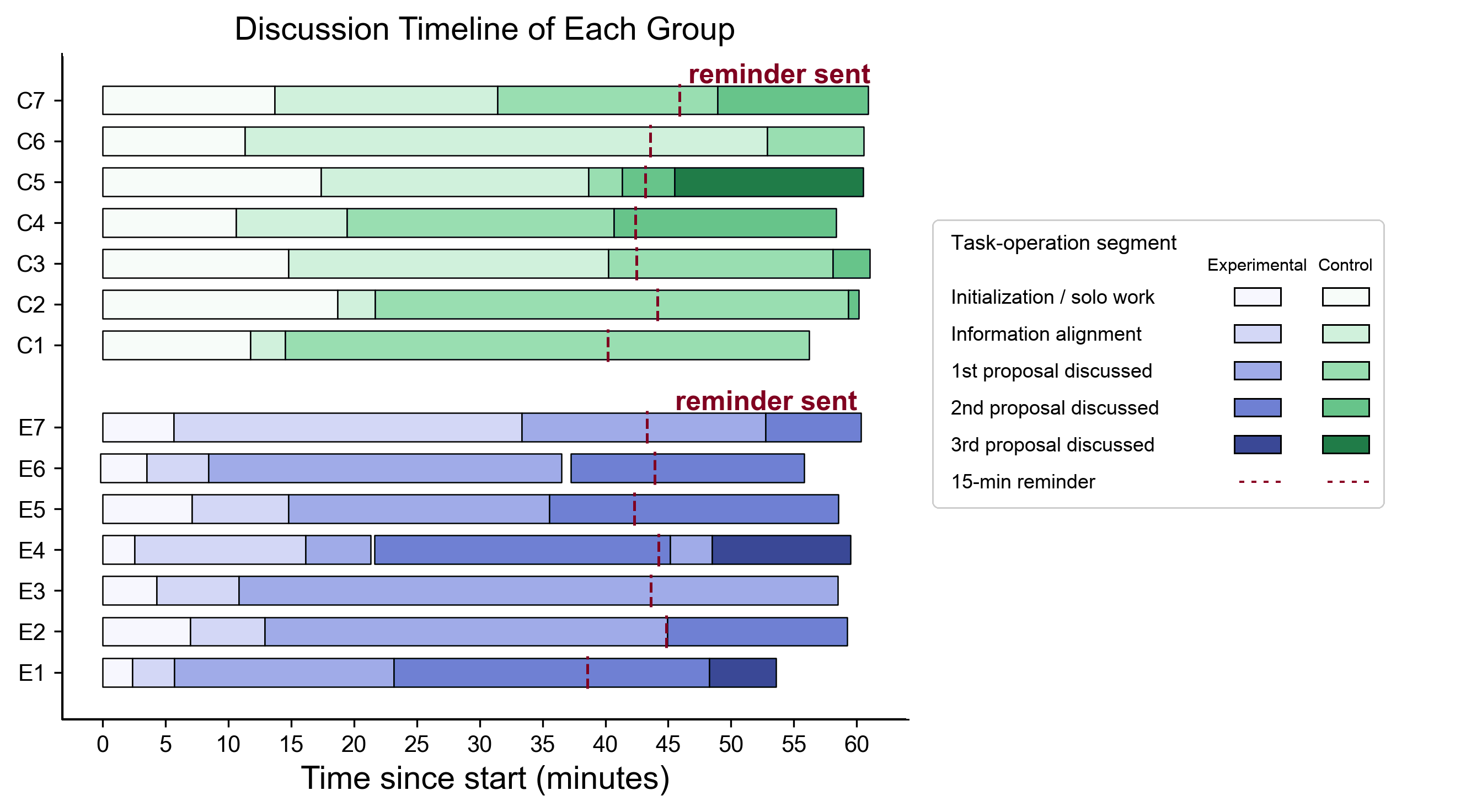}
        \caption{Task operation with each milestone and last-15-minute reminder alert marked.}
        \label{fig:task-operation}
    \end{subfigure}
    \captionsetup{skip=0.5ex}
    \caption{Task operation progress, including substage duration and milestone progress of experimental and control groups. Error bars depict standard deviations. The Welch's t-test (normal distribution) or Mann-Whitney U-test (otherwise) was employed for two groups. Significance values are reported for 0.05 < p < 0.1 ( $\hat{ }$ ), p < .05 (*), p < .01 (**), p < .001 (***), abbreviated by the number of stars. We calculated and presented Cohen's d (t-test) or rank-biserial r (U-test) as an indicator of effect size for significant comparisons.}
\end{figure}

\paragraph{\textbf{\textit{AnimaStand} Enabled Faster Transition from Initialization to Efficient Task Operation Substage, While Supporting More Thorough Discussion.}}
We first compared the temporal distribution of the three task-completion substages under the two conditions. As in Fig.~\ref{fig:sub-duration}, the control condition spent significantly more time in the \textit{initialization} substage ($5.53 \pm 3.33$ vs $14.05 \pm 3.10$ minutes), compressing the available time for the \textit{regular operation} substage ($37.45 \pm 3.90$ vs $29.07 \pm 2.98$ minutes). 
As in Fig.~\ref{fig:task-operation}, each group managed the discussion progress differently to complete the task. Experimental groups generally spent less time ($9.21 \pm 8.24$ vs. $17.50 \pm 13.39$ minutes) on information alignment before focusing on developing a specific proposal.
Given that experimental groups also discussed more proposals overall, this align-while-discussing strategy (used by 6/7 groups) appeared more time-efficient than the align-then-discuss approach favored in most control groups (4/7). 

\paragraph{\textbf{\textit{AnimaStand}'s Synchronous Movements Broke Group Silence and Ignited Teamwork Progress}}\label{silence-breaking}
Faster transition to the \textit{regular operation} substage and adoption of the align-while-discussing strategy may be thanks to the \textit{Silence‑Breaking} facilitation.
In the experimental condition, if all members were silent for over two minutes, phones stepped forward and rotated synchronously (\textit{Silence‑Breaking} in Fig.~\ref{fig:final-design}).
Across six experimental groups that experienced such silences in the \textit{initialization} substage, five groups initiated verbal discussions soon ($ 0.86 \pm 0.60$ min) after noticing phones' movements. In contrast, in the control groups, it took them an average of $14.05 \pm 3.10$ minutes until they could no longer endure this ``\textit{awkward and perplexed silence}'' (C2P1). 
Such external silence-breakers did more than just allocate limited session time toward productive dialogue. 
By breaking the prolonged period of silent individual reading typical of control groups, the early engagement emphasized the importance of interpersonal teamwork and prompted experimental groups to alter their workflow. 
Consequently, rather than attempting to align all information individually beforehand, they naturally adopted a more collaborative, iterative strategy (aligning information during coordinated task execution), which simultaneously fostered greater positive negotiation.
As C6P3 noted, ``\textit{It could be better if we had started aligning information earlier, then we may operate faster,}'' highlighting the benefits of earlier engagement.

\subsubsection{Post-Collaboration: Task Outcome Assessments}\label{task-overall-performance}
\begin{figure}[htbp]
    \centering
    \begin{subfigure}[t]{0.3\textwidth}
        \includegraphics[width=\linewidth]{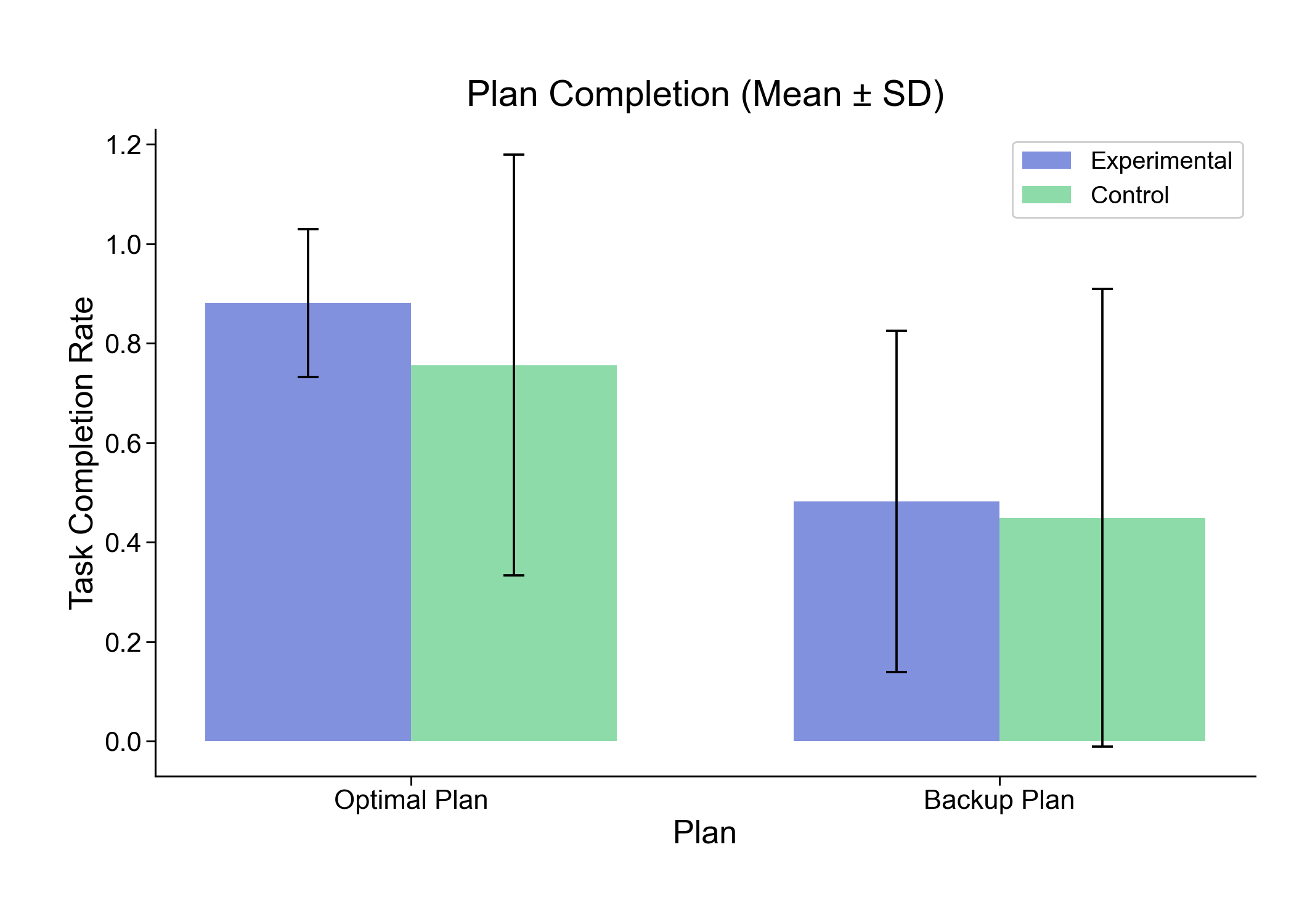}
        \captionsetup{skip=0.5ex}
        \caption{Task completion rate indicated by the completion ratio of the to-be-filled blanks in the plan form.}
        \label{fig:task-completion}
    \end{subfigure}
    \hspace{0.1\linewidth}
    \begin{subfigure}[t]{0.3\textwidth}
        \includegraphics[width=\linewidth]{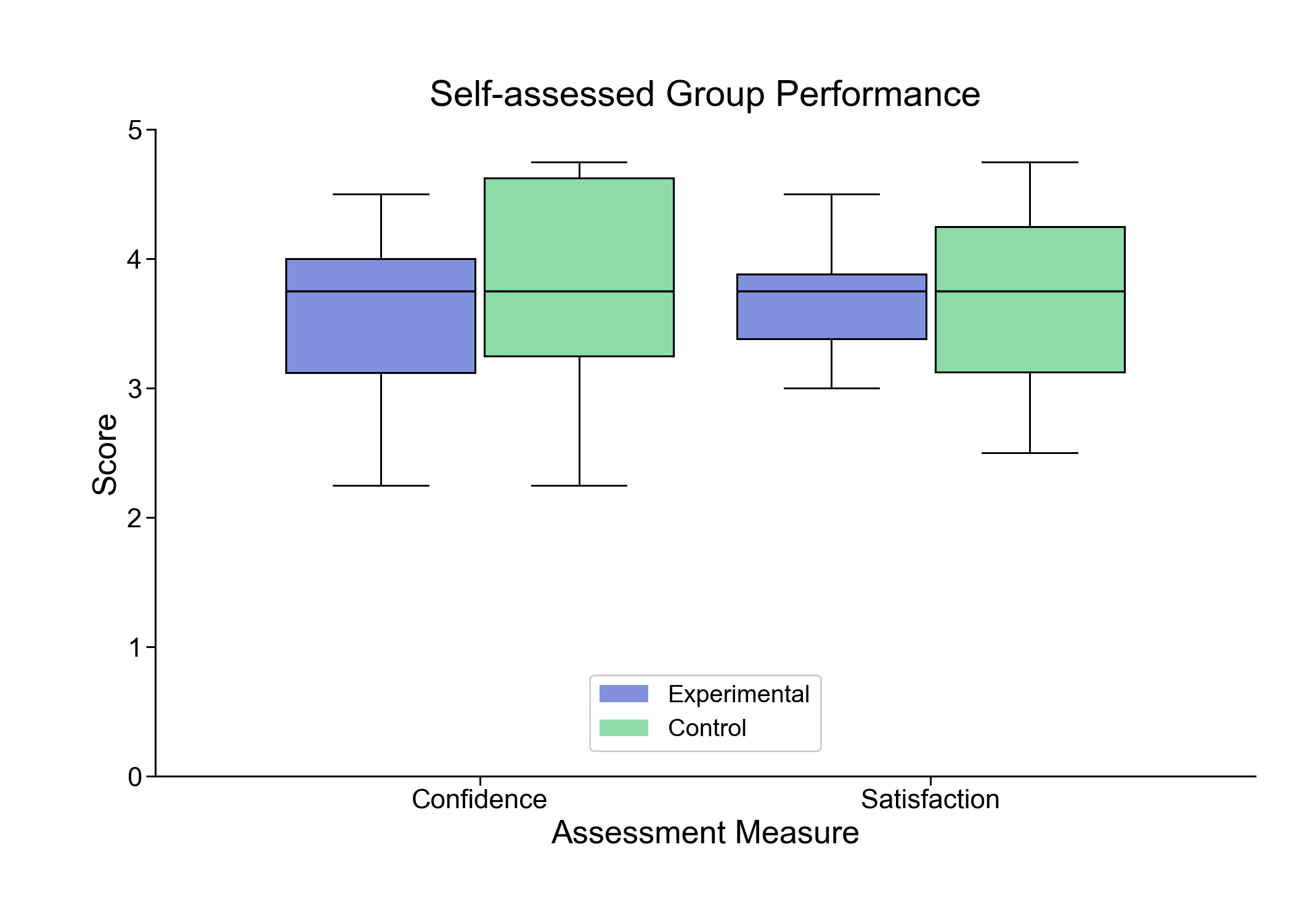}
        \captionsetup{skip=0.5ex}
        \caption{Self-assessed outcome presented in terms of Confidence and Satisfaction}
        \label{fig:outcome-assess}
    \end{subfigure}
    \captionsetup{skip=0.5ex}
    \caption{Task completion rate and self-assessed task outcome results. Error bars in the completion-rate panel indicate standard deviations; the self-assessment panel shows medians, interquartile ranges, and boxplot whiskers.}
\end{figure}

\paragraph{\textbf{\textit{AnimaStand} Showed No Significant Effect on Task Completion Performance}}
We then evaluated task completion outcomes objectively. Six of seven experimental groups delivered both the optimal and backup proposals within the allotted time, compared with four control groups; one control team produced no proposal. Quantitatively, experimental groups achieved completion rates of $88.1\% \pm 14.9\%$ for the optimal proposal and $48.2\% \pm 34.3\%$ for the backup, exceeding control group rates by $12.5\%$ and $3.3\%$, respectively, though these differences were not statistically significant. Similar to task completion rate, subjective self-reported confidence and satisfaction did not differ significantly between conditions (Fig.~\ref{fig:outcome-assess}).

\subsection{Peception and Interpretation of \textit{AnimaStand} (RQ3)}
\subsubsection{Perception of \textit{AnimaStand} (RQ3a)}
\label{perception}
\begin{figure}[ht]
  \centering
  \begin{subfigure}[]{0.35\textwidth}
    \includegraphics[width=\linewidth]{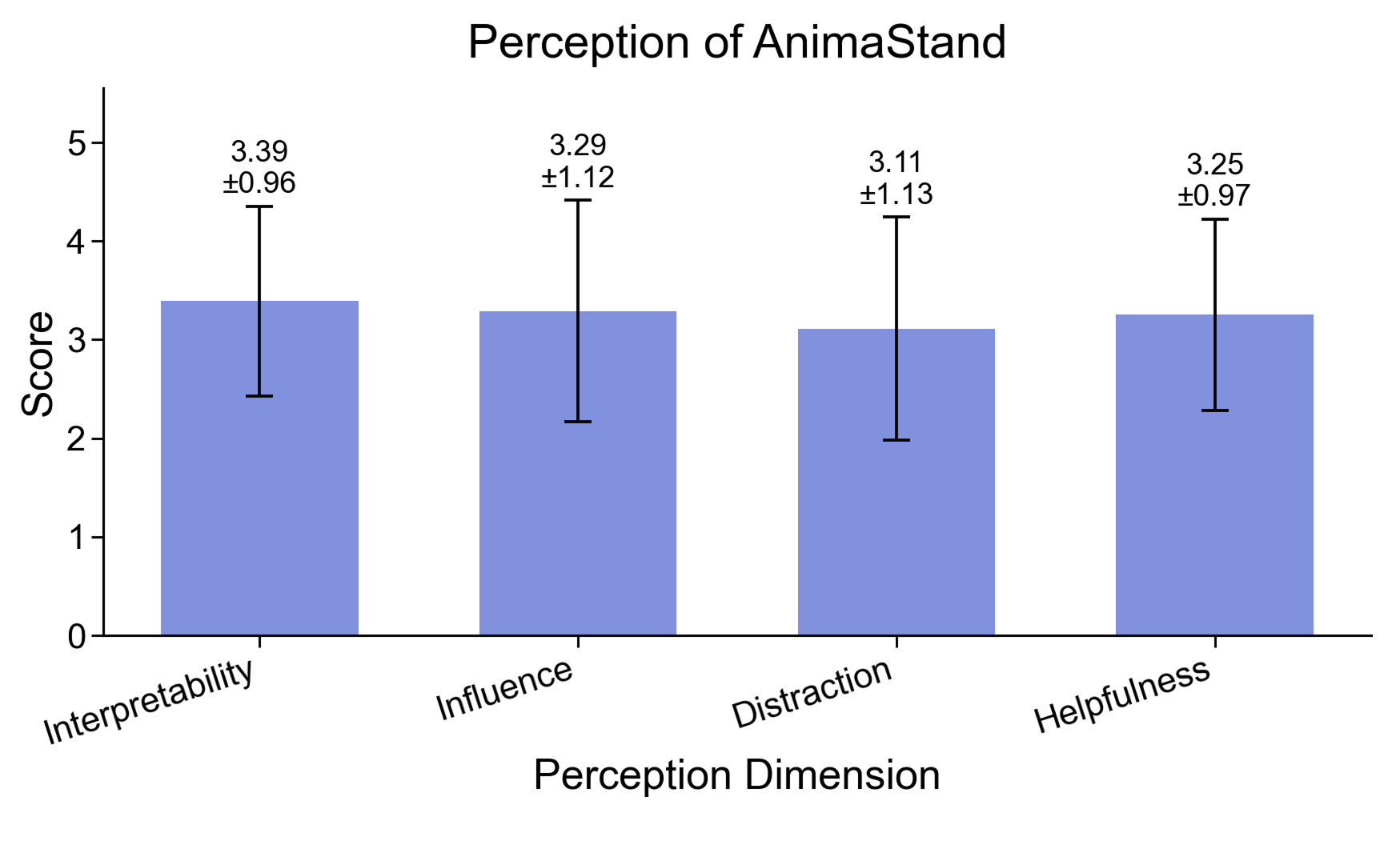}
    \label{fig:sub-first}
  \end{subfigure}
  \begin{subfigure}[]{0.35\textwidth}
    \includegraphics[width=\linewidth]{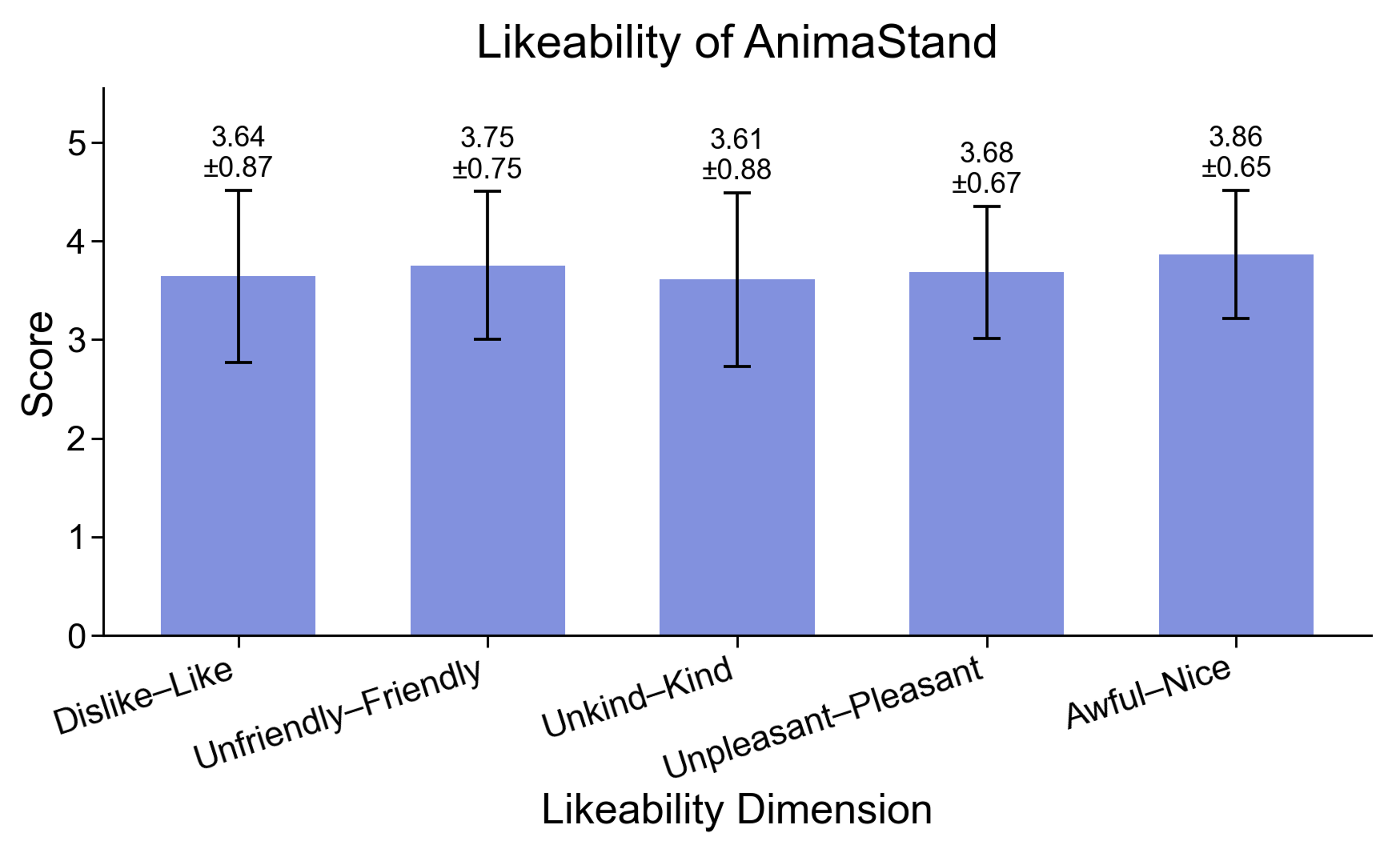}
    \label{fig:sub-last}
  \end{subfigure}
  \captionsetup{skip=0.5ex}
  \caption{Participants' Perception of the \textit{AnimaStand} in terms of Interpretability, Influence, Distraction, Helpfulness (left) and Likeability (right). Error bars depict standard deviations.}
  \label{fig:perception}
  \Description{}
\end{figure}

Besides the two dimensions focusing on facilitation effects, we finally present participants' perception of the \textit{AnimaStand} (Fig.~\ref{fig:perception}).
Overall, participants rated the \textit{AnimaStand} moderately high across all dimensions.
Particularly, for likeability-related ratings, all are (all means above 3.6 on a 5-point scale), with ``Awful–Nice'' receiving the highest mean of $3.86 \pm 0.65$, indicating generally positive affective impressions. As some participants (E1P3, E2P2, E4P2) echoed, ``\textit{It is fun to have the phone moving occasionally, while I can still use it.}'' 
However, although distraction received the lowest ratings, it still exceeded neutral, suggesting that distractions presented by \textit{AnimaStand} in group work require further attention.

Regarding expectations for future \textit{AnimaStand} development, personalization is a recurring theme. Participants hoped to customize \textit{AnimaStand}'s appearance (e.g., DIY phone cases or adding a display with owner information) (E7P2, E2P3) and movements to represent their personality better, echoing one participant's (E6P4) view that if they could design how phones move to ``\textit{represent myself,}'' they would be more willing to accept its cues.

\subsubsection{Interpretation of \textit{AnimaStand} (RQ3b)}
As reported above, participants were generally able to understand what the \textit{AnimaStand} meant. Next, we analyze how they interpreted its movements throughout the experiment.

\paragraph{\textbf{Participants Adopted Transferring Strategy when Interpreting Facilitation Movements}} 
The \textit{Storming} stage in our study was generally smooth in both conditions. This may be because participants were strangers and remained courteous despite a leader's bonus.
In the only two cases where \textit{Leader‑Election} facilitation was triggered (phones gathered, then pushed the member who had spoken most, with others blinking), the prompted participants both quickly (16s, 18s) guided the election process, though neither chose to become leader (E1P3, E5P4).
Such quick responses likely resulted from their recent exposure to facilitation in the \textit{Forming} stage, enabling them to transfer their understanding, ``\textit{I said something because from previous experiences I figured out when it comes out, I should speak}''(E5P4). 
This observation partially validates our design choice of modularizing phone behavior design, where shared movement patterns allow participants to reuse interpretive frames for faster, more effective responses.
% Contrary to common descriptions in the literature \cite{tuckman1977stages}, the \textit{Storming} stage, where participants elected roles, proceeded smoothly in both conditions. This may be because participants were strangers and remained courteous despite a leader's bonus.

\paragraph{\textbf{Participants Interpreted Facilitation Movements Through Prediction-and-Adaptation, Yielding Three Patterns of Interpretation Results}}\label{interpret-patterns}
Participants' sensemaking of \textit{AnimaStand} followed a typical pattern. They first \textbf{predicted} movement meanings using prior knowledge. For example, they linked phones facing others with displayed names to familiar name tag usage (E7P4).
Over time, participants may \textbf{adapt} their responsive behaviors iteratively according to \textit{AnimaStand}'s movements, as E3P1 reflected, ``\textit{When I hadn’t talked for a while, my phone moved out; when I spoke, it returned. After seeing this a few times, I knew I should speak when it moves out.}''
%Just like other sensemaking procedures, how participants construct interpretations of \textit{AnimaStand} follows a similar manner. 
%Participants usually first predict the meaning of the movements based on their knowledge, ``\textit{Though it took us a while to figure out what the phones are doing, when I connect the postures of our phones (facing others with a name on it) with the name tags we usually use in daily life, I'm easy to understand what it is expecting me to perform.}''(E7P4)
%Then, participants may adapt their responsive behaviors iteratively according to \textit{AnimaStand}'s movements over time, as E3P1 reflected, ``\textit{I noticed that when I hadn't talked for a while, my phone would move out; while when I started to speak, it would return and stay still. Having seen it move many times, I gradually know that I should speak when it moves out.}''

% Because this interpretation through prediction and adaptation is not anchored to any explicit ground truth, participants may map meanings for the facilitation movements that do not necessarily align with the original design intent. We identified three patterns of interpretation outcomes.
Because this predictive–adaptive interpretation is not anchored to an explicit ground truth, participants sometimes assigned meanings to facilitation movements that did not match the design intent. We observed three patterns of interpretation outcomes.
First, most interpretations \textbf{aligned with the design intent}, typically enabling expected responses. 
However, participants occasionally ignored prompts because they felt other matters were more important or because of cognitive or emotional fatigue. For example, ``\textit{When it prompts me to speak but I have nothing to say, I feel stressed and lose focus}'' (E4P2).
Second, some interpretations \textbf{diverged from the design intent}. 
In our design, we framed facilitation movements at a content-free behavioral level. They were intended to shape interactional dynamics such as switching speakers, managing speech duration, etc.
%In our design, we framed facilitation movements at a content-free behavioral level, intended to shape interactional dynamics (e.g., switching dominant contributors and managing speaking time) rather than specifying the content of the behavior (e.g., what to say).
Most ``misaligned'' interpretations instead mapped facilitation to broader task‑related behaviors with concrete content. For example, some participants interpreted the move‑out and rotate movements as ``share the hidden information'' rather than simply ``talk more.'' (E1P1, E6P4)
This misaligned content-based mapping sometimes still coincidentally led to behaviors that matched our intended effects: when participants were withholding information, they may respond by speaking (as we expected) and sharing it. However, when they were not hiding anything but were simply less engaged, this interpretation could lead to no action.
Finally, some \textbf{failed to form any clear interpretation}, e.g., ``\textit{I don't know how it operated, it is a bit disruptive}'' (E5P1). This results in puzzled gazes and minimal engagement, which is only a reaction toward a visual stimulus.

%% file: 7-discussion-new.tex
\section{Discussion}
Our findings show that animating personal smartphones can influence co-located group collaboration not merely by prompting individuals, but by making interactional issues publicly visible and inviting collective responses.
Interpersonal-wise, \textit{AnimaStand} re-engaged inactive members, caused more balanced participation and interdependence, and was associated with slightly higher perceived group oneness and equal peer recognition. Task-wise, it accelerated collaborative task operation yet did not produce significant improvements in objective task completion or subjective confidence and satisfaction. 
Participants perceived \textit{AnimaStand} as generally helpful despite occasional distraction, and usually interpreted its facilitations as intended, with some context-driven divergences.
Based on these findings, we first discuss the \textit{AnimaStand} facilitation mechanism in terms of engagement lift and interpretation. Drawing on these discussions, we further propose directions for developing adaptive and adaptable animated phone facilitation. Finally, we present the limitations and future works of this paper.
\subsection{Managing Engagement for Productive Group Collaboration}\label{config-optm-engage}
Our design of \textit{AnimaStand} implicitly assumed that higher and more equal engagement, particularly speech, as a pathway to supporting collaboration. 
In this section, we first examined the potential of this behavioral pathway for fostering overall engagement. 
Then, we discussed whether our assumption holds and what may constitute an optimal level of engagement for productive group collaboration, and how facilitation should be configured under different conditions to achieve it. 

\subsubsection{Potential of Boosting Behavioral Engagement to Lift Overall Engagement}\label{bhvengage-allengage}
Behavioral, emotional, and cognitive engagement are closely interrelated, and enhancing one may sometimes support the others. This raises a critical question: what forms of behavioral engagement may indicate deeper cognitive and emotional engagement with group work?
We observed that when participants responded by sharing opinions or consulting materials to answer others' questions, they actively committed attention to the collaborative task. Such \textit{\textbf{commitment}} may connect behavioral participation with deeper cognitive and emotional engagement. By contrast, low-effort follow-up questions, backchannels, or brief non-verbal responses often reflected \textit{\textbf{compliance}} rather than substantive contribution, sometimes introducing interactional ``noise'' (Sec.~\ref{noise}).
Although both commitment and compliance occurred in response to \textit{AnimaStand}'s cues, their divergence may depend on participants' momentary readiness and situational priorities. For example, participants might be fatigued, focused on individual work, or consider another activity more important at that moment. Therefore, fostering active \textit{\textbf{commitment}} rather than mere \textit{\textbf{compliance}} requires adaptive facilitation that accounts for participants' current readiness and priorities (discuss later in Sec.~\ref{adaptive}).

%\subsubsection{From Promoting Higher and More Equal Engagement of all members to Situational Facilitation tailored to each's role}
\subsubsection{From Equalizing Engagement to Context-Aware Situational Facilitation}
% 反思我们的方法，可能在两点上忽略了
% 忽略了group的主观能动性，从头到尾我们都一样facilitate，最后冲刺阶段experimental group也有两次facilitation。对于更成熟的团队，在他们的高速运转中不合时宜的facilitation可能是disruption，应该或许随着团队成熟，可以少一些facilitation
% 忽略了角色的不同，每个人的facilitation都是一样的。
Although experimental groups showed more active and balanced participation and turn-taking (Sec.~\ref{during-collaboration-interpersonal}), both objective and subjective task performance were not significantly higher (Sec.~\ref{task-overall-performance}). 
This indicates that while current animated facilitation, which emphasizes maximizing and equalizing interpersonal behavioral engagement, may reshape interaction toward more active and evenly distributed participation, it does not guarantee improved task outcomes.
%\lzq{The word not is missing? This should say does not guarantee; otherwise the meaning is reversed.}

Reflecting on our approach, the current facilitation strategy may not have sufficiently accounted for two contextual features of group collaboration, namely \textbf{group maturity} and \textbf{role criticality}.
First, our fixed triggering rules did not adapt to the group's growing capacity for self-regulation, aka.``maturity''. Even during the \textit{countdown operation} substage when experimental groups operated intensively, they still received approximately two facilitations. At this stage, additional facilitation may have become unnecessary or even disruptive. Therefore, following situational leadership principles \cite{weber1991student}, facilitation should gradually reduce its frequency and intensity as groups become more mature, preserving their autonomy and avoiding unnecessary disruption.
Second, the facilitation did not account for role criticality and applied equally to all members. Although balanced participation can support inclusiveness and shared ownership, but some tasks may benefit from temporary asymmetry when a particular role or capability becomes critical \cite{bligh2006importance, bass2014individual}. Facilitation applied to a critical member should therefore intensify when their contribution is missing, while fading when their role becomes less significant.

\begin{figure}[htbp]
    \centering
    \includegraphics[width=0.3\linewidth]{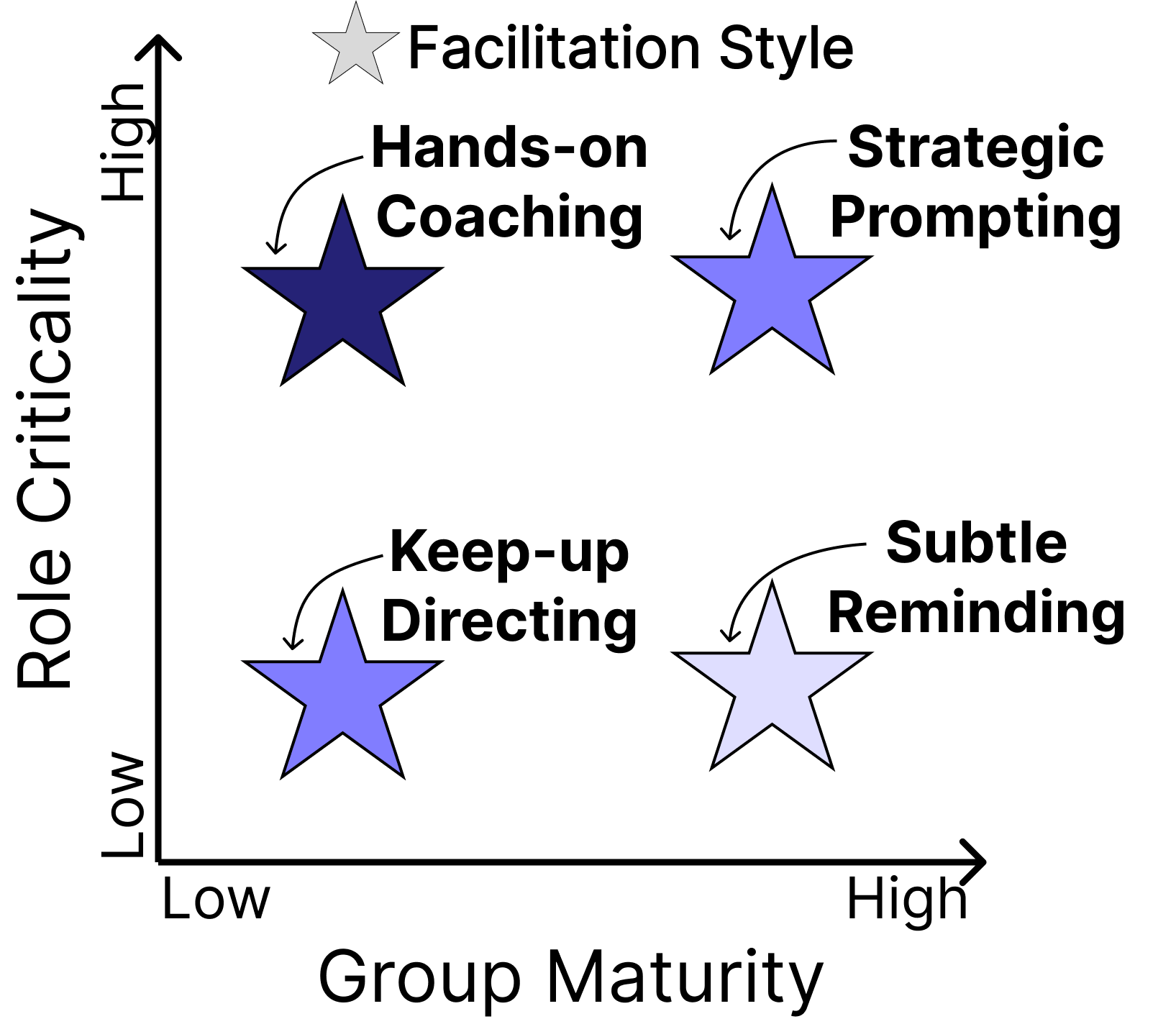}
    \captionsetup{skip=0.5ex}
    \caption{Role Criticality--Group Maturity Model, a situational configuration of animated facilitation.}
    \label{fig:config facilitate}
\end{figure}

Building on these perspective, we propose the \textit{Role Criticality--Group Maturity Model} in Fig.~\ref{fig:config facilitate} as a design lens for configuring animated facilitation.
Rather than prescribing fixed interventions, the model suggests adapting facilitation intensity and frequency along two dimensions: how critical a member is to the current task and how capable the group is of self-regulation.

\begin{comment}
\begin{figure}[htbp]
    \centering
    \begin{subfigure}[t]{0.3\textwidth}
        \centering
        \includegraphics[width=0.7\linewidth]{fig/ydlaw.png}
        \caption{Yerkes-Dodson's Law}
        \label{fig:ydlaw}
    \end{subfigure}
    \begin{subfigure}[t]{0.5\textwidth}
        \centering
        \includegraphics[width=0.5\linewidth]{fig/config-facilitate.png}
        \caption{Role Criticality--Group Maturity Model}
        \label{fig:config facilitate}
    \end{subfigure}
    \captionsetup{skip=0.5ex}
    \caption{Optimal engagement and situational configuration of animated facilitation.}
\end{figure}    
\end{comment}

Specifically, the model suggests four lightweight facilitation strategies:
\begin{itemize}
    \item \textbf{Hands-on Coaching} (high criticality, low maturity): frequent and salient cues to encourage a critical member's engagement.
    \item \textbf{Keep-up Directing} (low criticality, low maturity): frequent but less intense cues to keep a member engaged with group progress.
    \item \textbf{Strategic Prompting} (high criticality, high maturity): occasional salient cues at moments when a critical contribution is needed.
    \item \textbf{Subtle Reminding} (low criticality, high maturity): infrequent, low-intensity cues that preserve group autonomy.
\end{itemize}

We view this model as a design hypothesis rather than an empirically validated taxonomy.
Its value is to shift facilitation away from the assumption that every member should always participate equally, toward a more situational objective: providing the amount and form of intervention appropriate to the evolving group and task.

\subsection{Mechanisms and Pitfalls in Interpreting Animated Facilitation}
\subsubsection{Attention Shift toward Sensemaking the Ambiguous Animated Phone Facilitation}
Participants perceived \textit{AnimaStand} as both helpful and distracting (Sec.~\ref{perception}). We argue that this dual effect is rooted in the \textbf{ambiguity of animated movements}. Unlike an explicit textual instruction, a movement first attracts attention and then requires participants to infer why it occurred and how they should respond. This attention shift and sensemaking process can be either productive or disruptive.

On the positive side, momentary shifts in focus occasionally contributed to improved interpersonal relationships and diffused tension in conflicts, such as triggering shared laughter in awkward silence or redirecting quarrels toward commenting on the facilitation (Sec.~\ref{conflict}). 
However, it is questionable whether this is solely a benefit of facilitator design. It is also possible that positive feedback was affected by a novelty effect, which means that users tend to perceive new technology more positively in the first encounters \cite{smedegaard2019reframing, reimann2023social}. Longitudinal within‑participant studies are needed to separate novelty from sustained impact and to refine movement‑based cues that retain relational benefits without undermining task focus.

On the negative side, two distinct mechanisms may account for the perceived distraction accompanying sensemaking. 
First, some participants failed to interpret the movement or spent too long doing so, possibly because the current movements are relatively stiff and insufficiently interpretable, thus demanding more attentional resources. Second, even for those who understood the cues, the movement modality itself is attention‑grabbing, as also suggested by previous studies \cite{klein1976attention, oulasvirta2005interaction}.
To mitigate these distractions, several participants (E6P4, E3P3) suggested improving the transparency and interpretability of facilitation movements, for example, by providing concise pre‑task user guidance to reduce learning costs.
To reduce visual distractions from movements, motion designs with the slow-in-slow-out principle \cite{schulz2018classifying} and controlled saliency could be helpful.

\subsubsection{Misalignment Among Intended, Perceived, and Accepted Facilitation}
\paragraph{\textbf{Intended vs. Perceived}}
As shown in Sec.~\ref{interpret-patterns}, participants sometimes mapped movements to meanings beyond their design intent, often interpreting them as task-specific cues such as ``sharing hidden information.'' Such interpretations were not necessarily wrong, but suggest two design opportunities. First, current movements were largely presentational, behaving similarly toward owners and other members. More owner-directed pre-cues could clarify what the phone expects its owner to do, rather than merely what the phone itself appears to enact. Second, movements could communicate greater task relevance by varying features such as duration or repetition when a particular member's contribution becomes important.

\paragraph{\textbf{Perceived vs. Accepted}}\label{introvert}
Correct interpretation did not necessarily lead to acceptance (Sec.~\ref{interpret-patterns}). Participants sometimes resisted cues because of fatigue, competing priorities, or preferred interpersonal boundaries, similar to the compliance responses discussed in Sec.~\ref{bhvengage-allengage}. Although most participants were neutral or positive, E3P4 preferred to ``\textit{connect only with select members},'' while E2P3 described the phone as ``\textit{a shelter from overwhelming collaboration}.'' Both initially engaged with the social cues but later ignored or reluctantly followed them, showing that situational responses remain constrained by participants' relationship preferences.

\subsection{Directions for Future Development of Animated Facilitation for Group Collaboration: Adaptive and Adaptable}\label{directions}
\subsubsection{Adaptive to Individual and Group Contexts}\label{adaptive}
Our current facilitation is rule‑based and also relies solely on features observed from a third‑person perspective. 
While effective in tracking and guiding interaction flow, it overlooks temporal change in group context and divergent individual context.
We argue that a more efficient animated device facilitator should be more sensitive and adaptive to both contexts.
Building of previously discussed mechanisms, future facilitation should \textbf{adapt decisions (e.g., facilitation choice and form) to individual context, and adjust style (e.g., facilitation strategy) to group context}.

At the individual level, two contextual factors are particularly relevant: \textbf{relationship norms}, which shape how much interaction a participant desires, and \textbf{task priorities and situational awareness}, which determine their immediate focus and readiness to contribute. A system could initialize facilitation level based on lightweight user profiles and preferences, then update it using real-time signals from screen activity, cameras, and environmental sensors to infer participants' behaviors \cite{li2023human} and states \cite{frederiksen1999dynamic, hemminghaus2017towards}.

At the group level, the system should estimate current \textbf{role criticality} and \textbf{group maturity}. LLM-based analysis of task progress and discussion content could identify the current challenge and the member best positioned to contribute \cite{zhang2024imperative}. Combined with conversational dynamics and mental-model alignment within the group \cite{mathieu2000influence, van2011team}, it could also estimate the group's capacity for self-regulation. These estimates could then determine each member's facilitation style, operationalized through movement frequency, intensity, and triggering thresholds.

\subsubsection{Adaptable and Personalizable Animated Facilitation}
Personalization is another key direction proposed by participants. 
While phones are personal devices, the current design applies the same movements to all participants, not yet fully reflecting this connection. 
Thus, allowing users to customize their phone's movements could make the phone's behavior more similar to its owner's style and strengthen users' sense of alignment with the phone, which in turn may increase their willingness to accept facilitation (see participants' quotes in Sec.~\ref{perception}). 
However, personalized movements should be acceptable not only to the phone owner but also to other group members to effectively support collaboration, so the degree of personalization needs to be bounded. For each problematic situation, the underlying facilitation logic and basic movement pattern should remain consistent. For instance, when a group member is left out, that member's phone should always move forward to attract attention, regardless of who it is, while the specific way of moving forward (such as trajectory or moving style) can be customized.

\subsection{Limitation and Future Work}
This work has several limitations that inform directions for further exploration.
First, the study only involved 56 participants in 14 groups, statistical power of the results should be carefully interpreted.
Then, the current facilitation strategy and movement designs are rather rule-based and naive. We plan to implement the adaptive animated facilitation design as proposed in Sec.~\ref{directions} and evaluate its impact.
In addition, the current study focused on small-group facilitation within four-person groups and was conducted with short‑term interactions (about an hour); future work will examine longer‑term deployments in more complex groups to observe how animated facilitation influences sustained group development.
Moreover, we did not compare animated smartphones with generic moving objects, social robots, or other embodied facilitators. We therefore cannot determine whether the observed effects is closely related to smartphone's personal-device status or more generally from physical movements.
Finally, this work is an exploratory study, and there is a gap between our prototype's demonstrated facilitation potential and its acceptance as a real‑world product. We may develop more advanced stand-alone prototypes and focus more on user experience.

%% file: 8-conclusion.tex
\section{Conclusion}
This work explored how animated smartphones can facilitate small‑group in‑person collaboration. A design workshop identified facilitation needs across Tuckman's group stages, informing proxemics‑ and metaphor‑driven phone movements. We implemented these on \textit{AnimaStand}, a semi‑automated, movement‑enabled phone stand, and evaluated them in a between‑subjects Wizard‑of‑Oz study.
Results showed that animated facilitation improved engagement and balanced interactional dynamics by re‑engaging inactive members, leading to slightly elevated group relationships and equalized contributions. Task-wise, it accelerated collaborative task operation yet did not produce significant improvements regarding the task outcomes objectively and subjectively.
Ratings and interviews further indicated that \textit{AnimaStand} is generally found helpful, despite occasional distraction. Participants made sense of it through prediction‑and‑adaptation with a transferring strategy and yielded three interpretive patterns, highlighting that \textit{AnimaStand}'s effects emerged from the interplay between its publicly visible movements and participants' interpretations.
This work integrates group development, proxemics, and metaphor theories to design and study animated phone behaviors, showing how movement‑enabled devices can enhance group dynamics and relationships, informing future designs for adaptive and adaptable animated personal devices, and pathways for repurposing everyday devices.